\documentclass[fleqn,usenatbib]{mnras}

\usepackage{newtxtext,newtxmath}
\usepackage{chemformula}
\usepackage{booktabs}
\usepackage{pdflscape}

\usepackage[T1]{fontenc}

\usepackage{graphicx}	
\usepackage{amsmath}	
\usepackage{siunitx}
\usepackage[version=4]{mhchem}
\usepackage{xspace}
\usepackage{orcidlink}

\newcommand{\revision}[1]{#1}

\newcommand{\degree}{$^{\circ}$}

\newcommand{\mystar}{WASP-94A\xspace}
\newcommand{\myplanet}{WASP-94A\,b\xspace}
\newcommand{\eureka}{\texttt{Eureka!}\xspace}
\newcommand{\tiberius}{\texttt{Tiberius}\xspace}
\newcommand{\exoticjedi}{\texttt{ExoTiC-JEDI}\xspace}
\newcommand{\pRT}{\texttt{pRT}\xspace}
\newcommand{\hydra}{\texttt{HyDRA}\xspace}

\newcommand{\microns}{\textmu m\xspace}
\defcitealias{Carter2024AWASP-39b}{Carter \& May et al.\ 2024}
\defcitealias{Moran2023HighObservations}{Moran \& Stevenson et al.\ 2024}

\title[JWST NIRSpec transmission spectroscopy of WASP-94Ab]{Tracing the formation and migration history: molecular signatures in the atmosphere of misaligned hot Jupiter WASP-94Ab using JWST NIRSpec/G395H} 
\author[E. Ahrer et al.]{Eva-Maria Ahrer$^{\orcidlink{0000-0003-0973-8426},1}$\thanks{E-mail: ahrer@mpia.de},
Siddharth Gandhi$^{\orcidlink{0000-0001-9552-3709},2,3}$,
Lili Alderson$^{\orcidlink{0000-0001-8703-7751},4}$,
James Kirk$^{\orcidlink{0000-0002-4207-6615},5}$,
Johanna Teske$^{\orcidlink{0009-0008-2801-5040},6,7}$,\newauthor
Richard A. Booth$^{\orcidlink{0000-0002-0364-937X},8}$, 
Catriona H.~McDonald$^{\orcidlink{0000-0001-6442-6877},9}$, 
Duncan A. Christie$^{\orcidlink{0000-0002-4997-0847},1}$,
Alastair B. Claringbold$^{{\orcidlink{0000-0003-1309-5558}},2,3}$, \newauthor
Rebecca Nealon$^{\orcidlink{0000-0856-679X},2,3}$, 
Vatsal Panwar$^{\orcidlink{0000-0002-2513-4465},2,3}$ 
Dimitri Veras$^{\orcidlink{0000-0001-8014-6162},2,3, 10}$,
Hannah R.~Wakeford$^{{\orcidlink{0000-0003-4328-3867}},11}$
Peter J.~Wheatley$^{\orcidlink{0000-0003-1452-2240},2,3}$, \newauthor
Maria Zamyatina$^{{\orcidlink{0000-0002-9705-0535}},12}$
\\
$^{1}$Max-Planck-Institut f\"{u}r Astronomie, K\"{o}nigstuhl 17, 69117 Heidelberg, Germany\\
$^{2}$Centre for Exoplanets and Habitability, University of Warwick, Gibbet Hill Road, Coventry CV4 7AL, UK\\
$^{3}$Department of Physics, University of Warwick, Gibbet Hill Road, Coventry CV4 7AL, UK \\
$^{4}$Department of Astronomy, Cornell University, 122 Sciences Drive, Ithaca, NY 14853, USA\\
$^{5}$Department of Physics, Imperial College London, Prince Consort Road, SW7 2AZ, London, UK\\
$^{6}$Earth and Planets Laboratory, Carnegie Institution for Science, 5241 Broad Branch Road, NW, Washington, DC 20015, USA \\
$^{7}$The Observatories of the Carnegie Institution for Science, 813 Santa Barbara St., Pasadena, CA 91101, USA\\
$^{8}$School of Physics and Astronomy, University of Leeds, Leeds LS2 9JT, UK\\
$^{9}$Institute of Astronomy, University of Cambridge, Madingley Road, Cambridge, CB3 0HA, UK\\
$^{10}$Centre for Space Domain Awareness, University of Warwick, Gibbet Hill Road, Coventry CV4 7AL, UK\\
$^{11}$ School of Physics, University of Bristol, HH Wills Physics Laboratory, Tyndall Avenue, Bristol BS8 1TL, UK \\
$^{12}$Department of Physics and Astronomy, Faculty of Environment, Science and Economy, University of Exeter, Exeter EX4 4QL, UK\\
}

\date{Accepted XXX. Received YYY; in original form ZZZ}

\pubyear{\the\year{}}

\begin{document}
\label{firstpage}
\pagerange{\pageref{firstpage}--\pageref{lastpage}}
\maketitle

\begin{abstract}
The discovery of hot Jupiters that orbit very close to their host stars has long challenged traditional models of planetary formation and migration. Characterising their atmospheric composition --- mainly in the form of the carbon-to-oxygen (C/O) ratio and metallicity --- can provide insights into their formation locations and evolution pathways. With JWST we can characterise the atmospheres of these types of planets more precisely than previously possible, primarily because it allows us to determine both their atmospheric oxygen and carbon composition. 
Here, we present a JWST NIRSpec/G395H transmission spectrum from 2.8 -- 5.1 \microns of WASP-94Ab, an inflated hot Jupiter with a retrograde misaligned orbit around its F-type host star. We find a relatively cloud-free atmosphere, with absorption features of \ce{H2O} and \ce{CO2} at detection significances of $\sim 4\sigma$ and $\sim 11\sigma$, respectively. In addition, we detect tentative evidence of \ce{CO} absorption at $\sim3\sigma$, as well as hints of sulphur with the detection of \ce{H2S} at a $\sim 2.5\sigma$ confidence level. Our favoured equilibrium chemistry model determines a C/O ratio of $0.49^{+0.08}_{-0.13}$ for WASP-94Ab's atmosphere, which is substellar compared to the star's C/O ratio of $0.68 \pm 0.10$. The retrieved atmospheric metallicity is similar to the star's metallicity as both are $\sim 2\times$ solar. 
We find that this sub-stellar C/O ratio and stellar metallicity can be best explained by pebble accretion or planetesimal accretion in combination with large-distance migration of the planet.

\end{abstract}




\section{Introduction}

JWST has already started to revolutionise our understanding of exoplanet atmospheres, allowing us to study their composition in the infrared at high precision and explore new chemical processes, e.g., the detection of photochemically produced sulphur dioxide in the atmosphere of exoplanet WASP-39b \citep{Alderson2023EarlyG395H, Rustamkulov2023EarlyPRISM, Tsai2023PhotochemicallyWASP-39b}.  In transmission spectroscopy, precise measurements of the exoplanet transit depth as a function of wavelength are used to reveal the opacity sources acting in the planetary atmosphere \citep[e.g.,][]{Charbonneau2002DetectionAtmosphere}. Extending transmission spectra further into the infrared provides access to many additional molecules, in particular enabling observations of carbon species in exoplanet atmospheres. Hot Jupiters are ideal targets for atmospheric observations as they tend to have large atmospheric scale heights, due to high temperatures and large radii, resulting in strong features in transmission spectra. 

JWST transmission spectra have revealed a range of molecular species in the sample of hot Jupiters (\revision{here defined as exoplanets with } temperatures $1,000$\,K -- $2,000$\,K, radii R$_\mathrm{p} > 0.5$\,R$_\mathrm{Jup}$, masses M$_\mathrm{p}>0.2$\,M$_\mathrm{Jup}$) 
published thus far, demonstrating the diversity of hot Jupiter atmospheres. The detections range from carbon- and oxygen-bearing species, such as water vapour \citep[e.g.,][]{Ahrer2023EarlyNIRCam,Bell2023AWASP-43b,Feinstein2023EarlyNIRISS,Fu2024HydrogenExoplanet, Radica2023AwesomeNIRISS/SOSS,Taylor2023AwesomeObservations,Louie2024JWST-TSTSpectrum,Kirk2025WASP-15}, carbon dioxide \citep[e.g.,][]{Alderson2023EarlyG395H,Rustamkulov2023EarlyPRISM,TheJWSTTransitingExoplanetCommunityEarlyReleaseScienceTeam2023IdentificationAtmosphere,Xue20242}, carbon monoxide \citep[e.g.,][]{Esparza-Borges2023DetectionData,Grant2023DetectionG395H}, to sulphur species like sulphur dioxide \citep[e.g.,][]{Alderson2023EarlyG395H,Rustamkulov2023EarlyPRISM,Fu2024HydrogenExoplanet,Kirk2025WASP-15} and hydrogen sulphide \citep{Fu2024HydrogenExoplanet}. Further discoveries in hot Jupiter atmospheres include the detection of individual cloud species, silicate clouds, using the mid-infrared detector MIRI \citep[e.g.,][]{Grant2023JWST-TSTWASP-17b, Inglis2024QuartzB}.

Hot Jupiters are prime targets for atmospheric studies, while they are also important for research into planet formation and migration as both of these aspects are not fully understood. Leading planet formation theories such as core accretion \revision{\citep[e.g.,][]{Pollack1996FormationGas}} and gravitational instability \revision{\citep[e.g.,][]{Boss1997GiantInstability}} struggle to explain the formation of gas giants this close to their host stars \revision{where hot Jupiters are currently situated now \citep[e.g.,][]{Rafikov2005CanInstability,Rafikov2006AtmospheresInstability}: core accretion theory struggles as it cannot form a large enough core for run away gas accretion, and gravitationally unstable discs can only cool fast enough to fragment beyond several 10s of AU \citep[e.g., see review in ][]{Dawson2018OriginsJupiters}.}
Instead, they likely formed further out in the protoplanetary disc \revision{(via either formation pathway)} and migrated inwards, either via disc-driven migration or high-eccentricity migration \citep[e.g.,][]{Goldreich1980Disk-satelliteInteractions,Lin1996OrbitalLocation,Rasio1996DynamicalSystems,Weidenschilling1996GravitationalDistances}. 

To understand the dynamical history of hot Jupiters, we can investigate the distribution of alignments between the orbital planes of planets and the stellar spin axis of their host stars. For example, it has been suggested that low obliquities (near alignment of these two planes) point to disc-driven migration, while high obliquities point to high-eccentricity migration \citep[e.g.,][]{Fabrycky2009ExoplanetaryObservations}. However, there is evidence for tides damping obliquities after the initial stages of planet formation/migration, so low obliquity measurements do not necessarily mean disc-driven migration \citep[e.g.,][]{Albrecht2022StellarSystems}, in particular for older systems. Recent evidence also suggests that resonance locking could also drive damping of obliquities \citep[as well as orbital eccentricity and semi-major axis,][]{Zanazzi:2024}. Hence it is generally not possible to trace a planet's migration history simply from its measured obliquity unless the host star is located above the Kraft break \citep{Kraft1967:RotationBreak}. The Kraft break denotes the separation in rotational velocity between hotter stars (quickly rotating) and cooler stars with a thick convective envelope \citep[e.g.,][]{Dawson2014:HJTidal,Albrecht2022StellarSystems}. Tidal realignment is significantly more effective for cooler stars due to either the convective envelopes below the Kraft break \citep[e.g.,][]{Albrecht2012ObliquitiesMisalignments} or stellar gravity modes \citep{Zanazzi:2024}. Vice versa is also true: the host stars with radiative envelopes above the Kraft break are less likely to have realigned their hot Jupiter companions and as such they retain their primordial obliquities. Therefore, measured obliquities of hot Jupiters around hotter stars ($T_\mathrm{eff} \geq 6\,000$K) may allow a distinction between the two migration mechanisms to be made: misaligned orbits are caused by high-eccentricity migration and aligned orbits point to disc-driven migration \citep[e.g.,][]{Penzlin2024BOWIE-ALIGN:Compositions,Kirk2024BOWIE-ALIGN:History}.

It has also been suggested that the atmospheric compositions of hot Jupiters can give clues about where and how they formed \citep[e.g.,][]{Oberg2011TheAtmospheres,Madhusudhan2014TowardsMigration, Booth2017ChemicalDrift, Notsu2020TheDiscs, Schneider2021HowC/O, Penzlin2024BOWIE-ALIGN:Compositions}.
This approach is rooted in the fact that the composition of gas and solids in a protoplanetary disc varies radially as different volatiles occur in different states of matter due to temperature variations within the disc. 
Most recently, \citet{Penzlin2024BOWIE-ALIGN:Compositions} simulated the disc chemistry and migration of hot Jupiters and one of their main conclusions is that hot Jupiters with solar to super-solar metallicity have accreted substantial amounts of rocky material. In addition, their C/O ratios are dependent on the method of migration: hot gas giants that migrated within the disc are expected to have a lower C/O (accretion of inner disc material) compared to those that underwent high-eccentricity migration after disc dispersal \citep[see also ][]{Kirk2024BOWIE-ALIGN:History}. Interestingly, the relative C/O trend between misaligned and aligned planets may be reversed if silicate evaporation is taking place. Probing these predictions by characterising a sample of hot Jupiter atmospheres orbiting stars above the Kraft break is the primary goal of a larger, future study.  

Here we contribute an additional planet to this study, \myplanet, a hot Jupiter orbiting a star above the Kraft break in a misaligned, retrograde orbit. We present JWST NIRSpec/G395H observations of this planet, targeted by the JWST programme GO\,3154 (PI: Ahrer). 
First, we introduce the system in Section\,\ref{sec:wasp-94-system}, followed by a description of the observations and the data analysis in Section\,\ref{sec:data_reduction}. The atmospheric retrieval analysis is described in Section\,\ref{sec:retrievals} and we discuss our results in Section\,\ref{sec:discussion}. We conclude this manuscript with our findings in Section\,\ref{sec:conclusions}. Additional figures can be found in the Appendix Sections\revision{\,\ref{sec:appendix_catwoman-all-reductions}}, \ref{sec:appendix_priors}, \ref{sec:appendix_ret_results} and \ref{sec:appendix_corner_plots}.

\section{The WASP-94 system}
\label{sec:wasp-94-system}
\begin{figure}
    \centering
    \includegraphics[width=\linewidth]{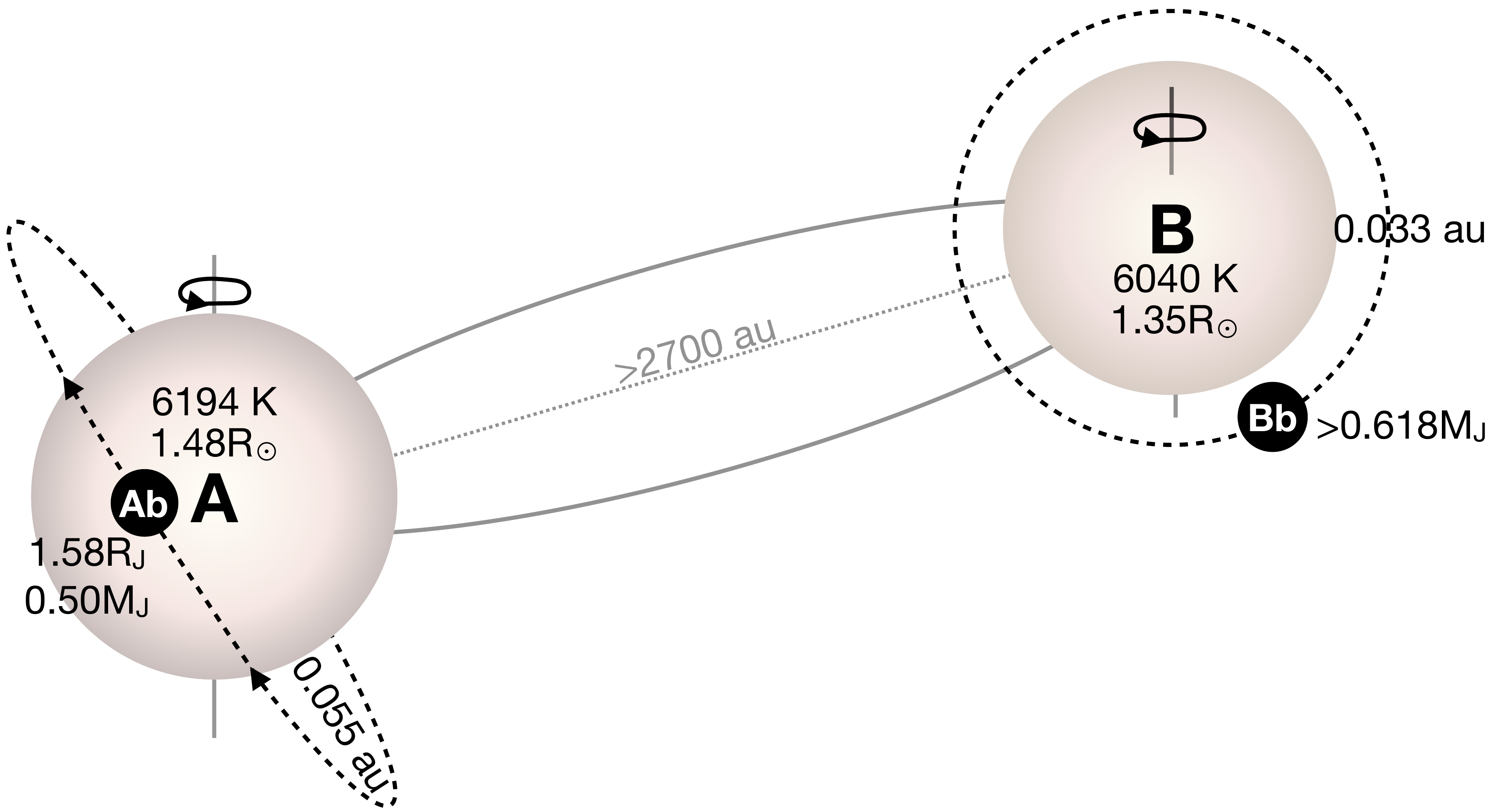}
    \caption{Sketch of the WASP-94 system. On the left, \myplanet is shown orbiting its host star at 0.056\,AU in a misaligned, retrograde orbit. On the right the non-transiting WASP-94B~b orbiting its host star at 0.033\,AU. Both stars are F type stars and are at least $>2700$\,AU apart. Only \mystar and \myplanet's radii ($R_\mathrm{p}/R_\mathrm{*}$) are to scale. }
    \label{fig:system-sketch}
\end{figure}

The WASP-94 system consists of two F-type stars, \mystar (F8) and WASP-94\,B (F9) with V magnitudes of 10.1 and 10.5, respectively. \revision{Their angular separation is $15.03\pm 0.01$~arcseconds with an orbital separation estimated to be $>2700$\,AU \citep{Neveu-Vanmalle2014WASP-94System}. One confirmed planet around each star \citep{Neveu-Vanmalle2014WASP-94System} was discovered as part of the Wide Angle Search
for Planets survey \citep[WASP,][]{Pollacco2006TheCameras}. Both planets orbiting the two stars in the WASP-94 system are hot Jupiters. \myplanet is a transiting exoplanet studied in this work, while WASP-94\,B\,b does not transit and has been detected by radial velocity measurements only. Therefore we cannot determine whether WASP-94\,B\,b is in a similar or different orbit to \myplanet or follow up with atmospheric studies.} 
The stellar parameters for \mystar are summarised in Table\,\ref{tab:wasp-94_parameters} \revision{and the system's architecture is sketched in Fig.\,\ref{fig:system-sketch}. }

\begin{table}

    \centering
    \caption{Stellar and planetary parameters for the \mystar planetary system. References are as follows [1] \citet{Hg2000TheStars}, [2] \citet[2MASS,][]{Skrutskie2006The2MASS}, [3] \citet{Neveu-Vanmalle2014WASP-94System},  [4] \citet{Teske2016THEB} from a differential analysis with WASP-94B, [5] \citet{Bonomo2017ThePlanets}, [6] \citet[Gaia DR3,][]{GaiaCollaboration2023GaiaProperties}, [7] \citet[Gaia DR3 GSP-Phot,][]{Andrae2023GaiaPhotometry}. }
    \begin{tabular}{l c c }
    \hline
   \multicolumn{3}{c}{\textbf{ Stellar parameters of \mystar}}\\
    \hline
        Parameter & Value &   Reference\\ \hline
        Brightness, $V_{\textrm{mag}}$ & $10.05 \pm 0.04$ & [1]\\
        Brightness, $J_\mathrm{mag}$ & $9.159 \pm 0.027$ & [2]\\
        Brightness, $K_\mathrm{s, mag}$ & $8.874\pm0.024$ & [2]\\
        Spectral type & F8  & [3]\\
        Effective Temperature, $T_{\textrm{eff}}$ (K) & $6194 \pm 5$ & [4]\\
        Age (Gyr) & $2.55 \pm 0.25$  & [4] \\
        Surface gravity, log $g$ (log$_{10}$(cm/s$^{2}$)) & $4.210 \pm 0.011$  & [4]\\
        Metallicity [Fe/H] (dex) & $0.320 \pm 0.004$ &  [4] \\
        Mass, M$_\textrm{*}$ (M$_\odot$) & $1.450 \pm 0.090$ & [5], [2]\\
        Radius, R$_\textrm{*}$ (R$_\odot$) & $1.5784^{+0.0095}_{-0.0110}$ & [6,7] \\  

        \hline
    \end{tabular}
    
    \label{tab:wasp-94_parameters}
\end{table}

\myplanet's radius was reported as $1.72^{+0.06}_{-0.05}$\,R$_\mathrm{Jup}$ \citep{Neveu-Vanmalle2014WASP-94System} and later refined as $1.58\pm0.13$\,R$_\mathrm{Jup}$ with the first Gaia parallaxes \citep{Stassun2017Parallaxes}. As the star's radius has been further refined with Gaia DR3 \citep[see Table\,\ref{tab:system_params};][]{GaiaCollaboration2023GaiaProperties,Andrae2023GaiaPhotometry} we adopt the planetary radius calculated by the weighted mean of our transit depth measurements multiplied by the DR3 stellar radius in our atmospheric retrieval studies. The mass of \myplanet was found as $0.452^{+0.035}_{-0.032}$\,M$_\mathrm{Jup}$ \citep{Neveu-Vanmalle2014WASP-94System} and the orbit is consistent with zero eccentricity \citep{Bonomo2017ThePlanets}. An equilibrium temperature of $1508 \pm 75$~K was derived using \textit{Spitzer} eclipses \citep{Garhart2020Eclipses}, which were also consistent with the previously derived circular orbit. Most interestingly, \myplanet is in a retrograde, misaligned orbit around its host star, tightly constrained with recent HARPS measurements of its projected obliquity $\lambda = -123.0 \pm 3.0$ \citep{Ahrer2024AtmosphericHARPS}. 
We discuss \revision{the observed orbit of \myplanet and its implications for planet formation and migration within a stellar binary} in Section\,\ref{sec:implications-planet-formation-dynamics}.  


In \cite{Teske2016THEB}, the authors took advantage of the ``twin'' nature of WASP-94A and B to conduct a stellar abundance analysis looking for compositional anomalies that might indicate differences in how planet formation proceeded around each star, which is suggested by the retrograde and misaligned orbit of \myplanet. Based on their high-precision, strictly differential abundance analysis, \cite{Teske2016THEB} found a slight depletion of volatile elements ($\sim$-0.02 dex on average) but enhancement of refractory elements ($\sim$0.01 dex) in WASP-94\,A versus B. Whether or not these anomalies between the twin stars are actually related to planet formation is unclear. 
We revisit the abundance measurements of WASP-94\,A in Section\,\ref{sec:discussion}.

\myplanet's atmosphere has been previously studied at low spectral resolution with the EFOSC2 spectrograph on the New Technology Telescope (NTT) as part of the Low Resolution Ground-Based Exoplanet Atmosphere Survey using Transmission Spectroscopy (LRG-BEASTS), as well as at high spectral resolution with the HARPS spectrograph \citep[High Accuracy Radial Velocity Planet Searcher,][]{Mayor2003SettingHARPS}, both located at La Silla, Chile. \citet{Ahrer2022LRG-BEASTS:NTT/EFOSC2} presented a low resolution transmission spectrum of \myplanet from 4020 -- 7140\,\AA, showing evidence for a scattering slope and detecting Na absorption. \citet{Ahrer2024AtmosphericHARPS} found comparable Na signatures using the high spectral resolution of HARPS by resolving the Na doublet in the in-transit observations of \myplanet.

\section{Data reduction and analysis}
\label{sec:data_reduction}
We analyse our new JWST NIRSpec/G395H observations using three independent reductions (\eureka, \tiberius, \exoticjedi) to ensure our results are robust against reduction choices. We outline these in this section after a description of our observational setup.

\subsection{Observations}
\label{sec:observations}
Our JWST observations of \myplanet (GO 3154, PI:Ahrer) took place on 7 June 2024 using NIRSpec's highest resolution grism G395H/F290LP. We used 21 groups per integration, NRSRAPID readout mode, with 1451 total integrations and an overall exposure time of 8\,hours. The visit covered a total time of 2.25\,hours of pre-transit time, a transit time of $4.5$\,hours (including ingress and egress) and a post-transit time of 1.25\,hours. Target acquisition was done using a fainter star (2MASS J20550897-3408314) within the splitting distance, using SUB32 subarray, 3 groups/integration, F140X filter and NRSRAPID readout.

\subsection{\eureka}
We reduced the \myplanet's observations using the open-source \texttt{Python} package \eureka \citep{Bell2022Eureka:Observations} which has been successfully applied to a multitude of observations \citep[e.g.,][]{Ahrer2023EarlyNIRCam,Bell2023MethaneWASP-80b,Fu2024HydrogenExoplanet}.

\subsubsection{Light curve extraction}
To extract the time-series stellar spectra, we started with the uncalibrated \texttt{uncal.fits} files and ran \eureka's Stage\,1\&2 which are wrapped around the default \texttt{jwst} pipeline. We followed the default \texttt{jwst} steps in both stages, with the exception of the jump step where we used a threshold of 10$\sigma$ (instead of the default 4$\sigma$) and we skipped the \texttt{photom\_step}. In addition, we applied \texttt{Eureka!}'s group-level background subtraction to reduce the 1/f noise by subtracting a zero-order polynomial after masking the trace and outliers $>5$ times the median. A custom bias scale factor was also used by computing a smoothing filter \citepalias[see also][]{Moran2023HighObservations}.

We extracted the stellar time-series spectra in \eureka's Stage\,3. First, we performed a trace curvature correction, followed by a column-by-column background subtraction. For the background subtraction we used a zero-order polynomial fitted to each column, excluding the area within 10 pixels of the centre of the trace and masking outliers with thresholds of 5$\sigma$ along the time and the spatial axis. Finally, we extracted the time-series spectra using an aperture size of 9 pixels. 

We used Stage\,4 of \eureka to generate binned spectroscopic light curves of our observations, as well as broad white-light light curves of the full NRS1 (2.8 -- 3.7 micron) and NRS2 (3.8 -- 5.2 micron) wavelength ranges. Prior to Stage\,4, we manually masked one bad wavelength column in NRS2 and one for NRS1 which we found were not masked during Stage\,3 and introduced outliers in the light curves and subsequently in the transmission spectrum. In \eureka's Stage\,4 we utilised a 5$\sigma$-clipping with a 20-pixel-rolling-median to mask outliers in the light curves. 

\revision{For all our reductions, we use two different binning schemes when generating our light curves, at a resolution of R=100 and R=400. This allows us to test our atmospheric inferences at different resolutions and determine whether there are outlier spectral channels that are binned over in the lower resolution data, while also resolving features in the planet's atmosphere. This follows the recommended strategy by the in-depth study of WASP-39b using JWST's NIRSpec/G395H, NIRSpec/PRISM, NIRCam/F322W2 and NIRISS/SOSS \citepalias{Carter2024AWASP-39b} and is commonly used in other JWST observations of hot Jupiters \citep[e.g.,][]{Kirk2025WASP-15,Meech2025BOWIE-ALIGN:Spectroscopy}}.

\subsubsection{Light curve fitting}

\begin{figure}
    \centering
    \includegraphics[width=\columnwidth]{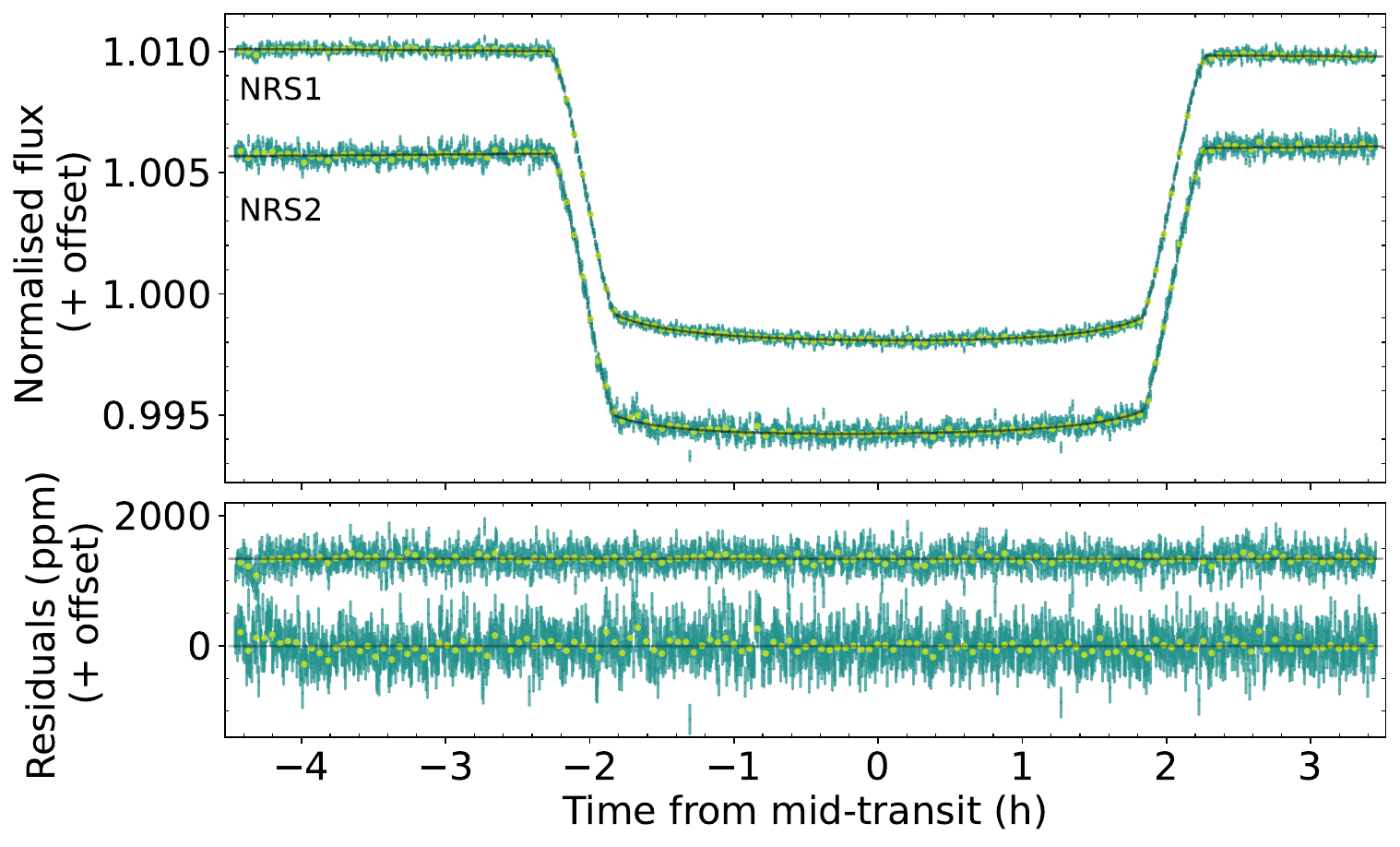}
    \caption{Top: Broad-band light curves of the transit of \myplanet for the two NIRSpec detectors, NRS1 and NRS2, using the \eureka reduction. The black line denotes the fit to the data, while the light and darker green colours represent the unbinned and binned data (10-pixel binning). Bottom: Respective residuals of the fit to the light curves from the top panel.  }
    \label{fig:WL_curves}
\end{figure}

\eureka's Stage\,5 performs the light curve fitting. First, we fit the white light curve for both NRS1 and NRS2 separately, freely fitting for the orbital parameters time of mid-transit (T$_0$), scaled semi-major axis ($a/R_*$), inclination ($i$), transit depth ($R_\mathrm{p}$/$R_*$) as well as two parameters to fit a linear trend and one limb-darkening parameter ($u2$). We used the quadratic limb-darkening law and fixed the second parameter, $u1$, to the value calculated using the \texttt{ExoTIC-LD} tool \citep{Grant2024ExoTiC-LD:Coefficients} and a 3D stellar atmosphere grid \citep{Magic2015TheCoefficients} based on the stellar parameters in Table\,\ref{tab:wasp-94_parameters}. We fix the orbital period of \myplanet to 3.9502001\,days \citep{Kokori2023ExoClockObservations}. 

For all light curve fitting, we utilise the \texttt{batman} package for our transit model and the Markov Chain Monte Carlo (MCMC) package \texttt{emcee} \citep{Foreman-Mackey2013EmceeHammer} to retrieve our fitted parameters. The fitted values from the white light curve for NRS1 and NRS2 are summarised in Table\,\ref{tab:system_params}. 

\begin{figure}
    \includegraphics[width=\linewidth]{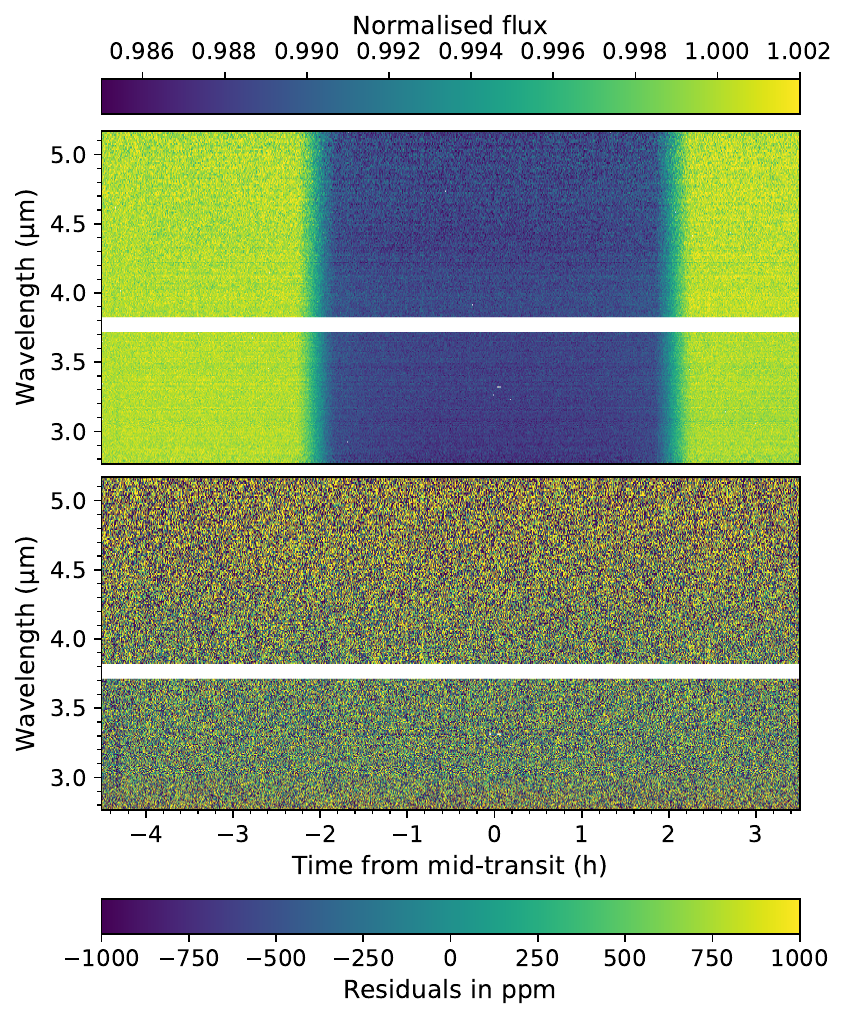}
    \caption{Light curves of \myplanet's transit and corresponding residuals using \eureka, at $R=400$ for both NRS1 and NRS2 detectors. }
    \label{fig:spectroscopic_light_curves}
\end{figure}

For the spectroscopic light curve fits we fixed the parameters $i$, $a/R_*$ and T$_0$ to the retrieved values from the white light curve fits. Therefore only $R_\mathrm{p}$/$R_*$, $u2$ and the two linear trend parameters are freely fitted for each spectroscopic light curve. \revision{The spectroscopic light curves at a spectral resolution of R=400 are shown in Fig.\,\ref{fig:spectroscopic_light_curves}.}

\begin{table*}
    \caption{The retrieved system parameters for the transit of \myplanet from the JWST NIRSpec/G395H white light curves as fitted by the individual reductions. }
    \label{tab:system_params}
    \centering
    \begin{tabular}{cccccc} \hline
         Pipeline & Detector & $T_0$ (BJD) & $R_\mathrm{p}/R_*$ & $a/R_*$ & $i$ (\degree) \\ \hline 
         \eureka & NRS1 & $24560469.308647 \pm 0.000019$ & $0.10608 \pm 0.00011$ & $7.235 \pm 0.018$ &  $88.42 \pm 0.11$ \\ 
         \eureka & NRS2 & $24560469.308671 \pm 0.000033$ & $0.10560 \pm 0.00060$ & $7.279 \pm 0.031$ &  $88.70 \pm 0.24$ \\ 
         \eureka & Weighted mean & $24560469.308656 \pm 0.000017$ & $0.10606 \pm 0.00011$ & $7.246 \pm 0.016$ & $88.47 \pm 0.10$ \\ \hline
         \tiberius & NRS1 & $24560469.308645 \pm 0.000019$  & $0.106022 \pm 0.000039$ & $7.252 \pm 0.016$ & $88.54 \pm 0.09$  \\ 
         \tiberius & NRS2 & $24560469.308666 \pm 0.000028$ & $0.105583 \pm 0.000058$ & $7.284 \pm 0.024$ & $88.71 \pm 0.16$ \\ 
         \tiberius & Weighted mean & $24560469.308652 \pm 0.000016$ &  $0.105885 \pm 0.000032$ & $7.262 \pm 0.013$  & $88.58 \pm 0.08$  \\ \hline
         \exoticjedi & NRS1 & 2460469.308631 $\pm$ 0.000019 & 0.105798 $\pm$ 0.000039 & 7.255 $\pm$ 0.016 & 88.57 $\pm$ 0.10 \\ 
         \exoticjedi & NRS2 & 2460469.308651 $\pm$ 0.000027 & 0.105502 $\pm$ 0.000055 & 7.276 $\pm$ 0.022 & 88.67 $\pm$ 0.15 \\ 
         \exoticjedi & Weighted mean & $2460469.308638 \pm 0.000016$ & $0.105699 \pm 0.000032$ & $7.262 \pm 0.013$  & $88.601 \pm 0.084$  \\ \hline
    \end{tabular}
\end{table*}

\subsection{\tiberius}

For our second independent reduction of the data, we used \texttt{Tiberius} \citep{Kirk2017RayleighHAT-P-18b,Kirk2021ACCESSWASP-103b} that has been used in several JWST analyses to date \citep[e.g.,][]{TheJWSTTransitingExoplanetCommunityEarlyReleaseScienceTeam2023IdentificationAtmosphere,Rustamkulov2023EarlyPRISM,Kirk2024JWST/NIRCam341b,Meech2025BOWIE-ALIGN:Spectroscopy}. 

\subsubsection{Light curve extraction}

Our light curve extraction proceeded in an identical way to that presented in \citet{Kirk2025WASP-15}. In brief, we process the \texttt{uncal.fits} files through stage 1 of the \texttt{jwst} pipeline and then feed the resulting \texttt{gainscalestep.fits} files into \tiberius. The only difference between \cite{Kirk2025WASP-15} and our application of \tiberius here is that we made a fresh bad pixel mask. This was motivated by potential changes in pixel behaviour between the different observation epochs. While we use the same reference files as \cite{Kirk2025WASP-15}, our custom step to flag outliers has the ability to identify pixels that would be outliers in one observation and not another. 

\subsubsection{Light curve fitting}

For our light curve fitting, we adopted the same procedure as for the analysis of WASP-15b \citep{Kirk2025WASP-15}. Specifically, our light curve model consisted of an analytic \texttt{batman} \citep{Kreidberg2015BatmanPython} transit light curve multiplied by a linear-in-time polynomial. The parameter space was explored with a Levenberg-Marquadt algorithm within the \texttt{scipy} library \citep{scipy}. Our fits were performed in two iterations, with the first iteration used to rescale the photometric uncertainties by a factor of 1.4 to give $\chi^2_{\nu} = 1$. The second iteration was used to infer the best-fit parameters and uncertainties.

We began by fitting the white light curves to derive a common set of system parameters. Similarly to the \eureka reduction, the free parameters here were the time of mid-transit ($T_0$), the scaled semi-major axis ($a/R_*$), the planet's orbital inclination ($i$), the scaled planet radius ($R_\mathrm{p}/R_*$) and the two coefficients of the linear polynomial. We fixed the period to 3.9502001\,d \citep{Kokori2023ExoClockObservations}. 
Similar to the \eureka reduction, we parameterised the limb darkening with a quadratic law though with both limb darkening coefficients fixed to values computed using \texttt{ExoTIC-LD} \citep{Grant2024ExoTiC-LD:Coefficients}, 3D stellar atmosphere models \citep{Magic2015TheCoefficients} and the stellar parameters of \citep{Bonomo2017ThePlanets}.

The best-fit system parameters are given in Table \ref{tab:system_params} and we find these are consistent between the detectors. We then fitted the spectroscopic light curves with $a/R_*$, $i$ and $T_0$ fixed to the mean-weighted values from our white light fits. This meant that only $R_\mathrm{p}/R_*$ and the two coefficients of the linear polynomial were fit parameters. 

We find minimal red noise in both the fits to the white and spectroscopic light curves which allows us to obtain a precise transmission spectrum, with a median uncertainty of 37\,ppm and 71\,ppm at $R=100$ and $R=400$, respectively.

\subsection{\exoticjedi}
Our third independent reduction uses \exoticjedi \citep{Alderson2022Exo-TiC/ExoTiC-JEDI:V0.1-beta-release}. Our reduction followed the same process as previous \exoticjedi reductions of other datasets \citep[e.g.,][]{Alderson2023EarlyG395H, May2023DoubleG395H, Alderson2024JWSTTOI-836b, Alderson2025JWSTTOI-776b}, treating NRS1 and NRS2 separately.

\label{sec:exotic-jedi}
\subsubsection{Light curve extraction}
We begin with the \texttt{uncal} files in Stage 1, a modified version of the \texttt{jwst} pipeline (v1.14.0, \citealt{Bushouse2024JWSTPipeline}), performing linearity, dark current, saturation and ramp jump corrections (with a threshold of 15$\sigma$, as opposed to the default 4$\sigma$), the \exoticjedi custom destriping routine to remove 1/$f$ noise at the group level, the \exoticjedi custom bias subtraction, and finally ramp fitting. In Stage 2, we performed the standard temporal and spatial pixel outlier cleaning and further 1/$f$ and background removal. To extract the 1D stellar spectra we used an aperture region five times the Full Width Half Maximum (FWHM) of Gaussians fitted to each column of the spectral trace, equivalent to approximately 7 pixels from edge to edge. We additionally cross-correlated the resulting spectra to obtain $x$- and $y$-pixel positional shifts to be used as systematic detrending parameters in our light curve fits.

\subsubsection{Light curve fitting}

We fitted white light curves for NRS1 and NRS2 across the G395H wavelength range (2.814--3.717\micron\ and 3.824--5.111\micron\ respectively), which were used to inform the spectroscopic light curve fits. For the white light curves, we fitted for $R_\mathrm{p}/R_*$, $i$, $a/R_*$ and $T_0$, holding the period fixed to the value presented in \citet{Neveu-Vanmalle2014WASP-94System}. We calculated and held fixed stellar limb-darkening coefficients using \texttt{ExoTiC-LD} \citep{Grant2024ExoTiC-LD:Coefficients}, with the non-linear limb-darkening law \citep{Claret2000AGravities} using the \citet{Magic2015TheCoefficients} 3D stellar models and stellar parameters from \citet{Stassun2017Parallaxes}. We used a least-squares optimiser to fit for a \citet{Kreidberg2015BatmanPython} transit model simultaneously with a systematic model $S(\lambda)$ of the form
$$ S(\lambda) = s_0 + (s_1 \times t) + (s_2 \times x_{s}|y_{s}|) \mathrm{,}$$
where $x_s$ is the $x$-positional shift of the spectral trace, $|y_s|$ is the absolute magnitude of the $y$-positional shift of the spectral trace, $t$ is the time and $s_0, s_1, s_2$ are coefficient terms. During the fitting process, we removed any data points that were greater than 4$\sigma$ outliers in the residuals and refitted the light curves until no such points remained. We additionally rescaled the light curve errors using the beta value \citep{Pont2006TheDetection} to account for any remaining red noise in the data.

\begin{figure*}
    \centering
    \begin{minipage}{0.495\textwidth}
        \includegraphics[width=\textwidth]{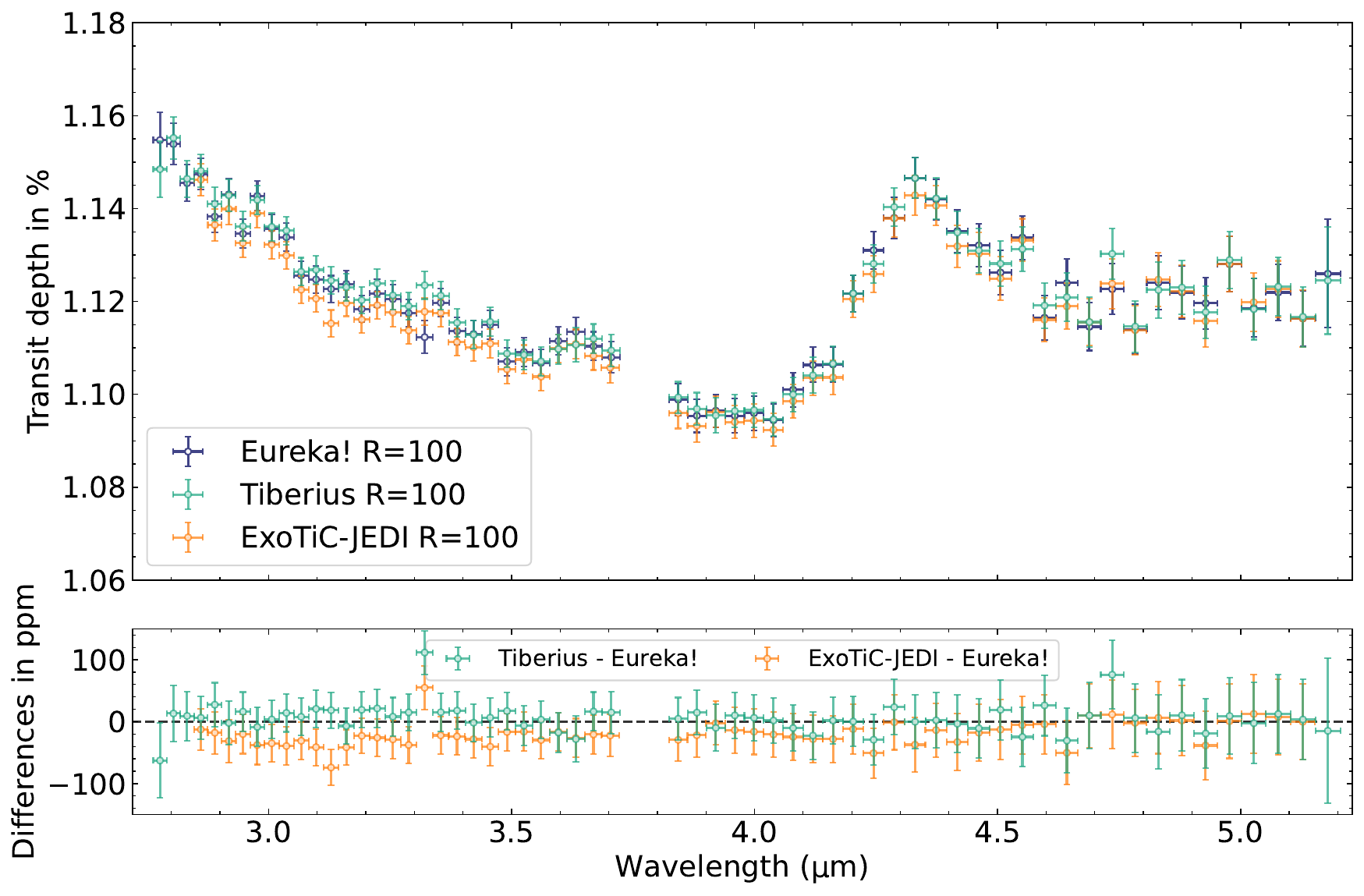}
    \end{minipage}
    \hfill
    \begin{minipage}{0.495\textwidth}
    \includegraphics[width=\textwidth]{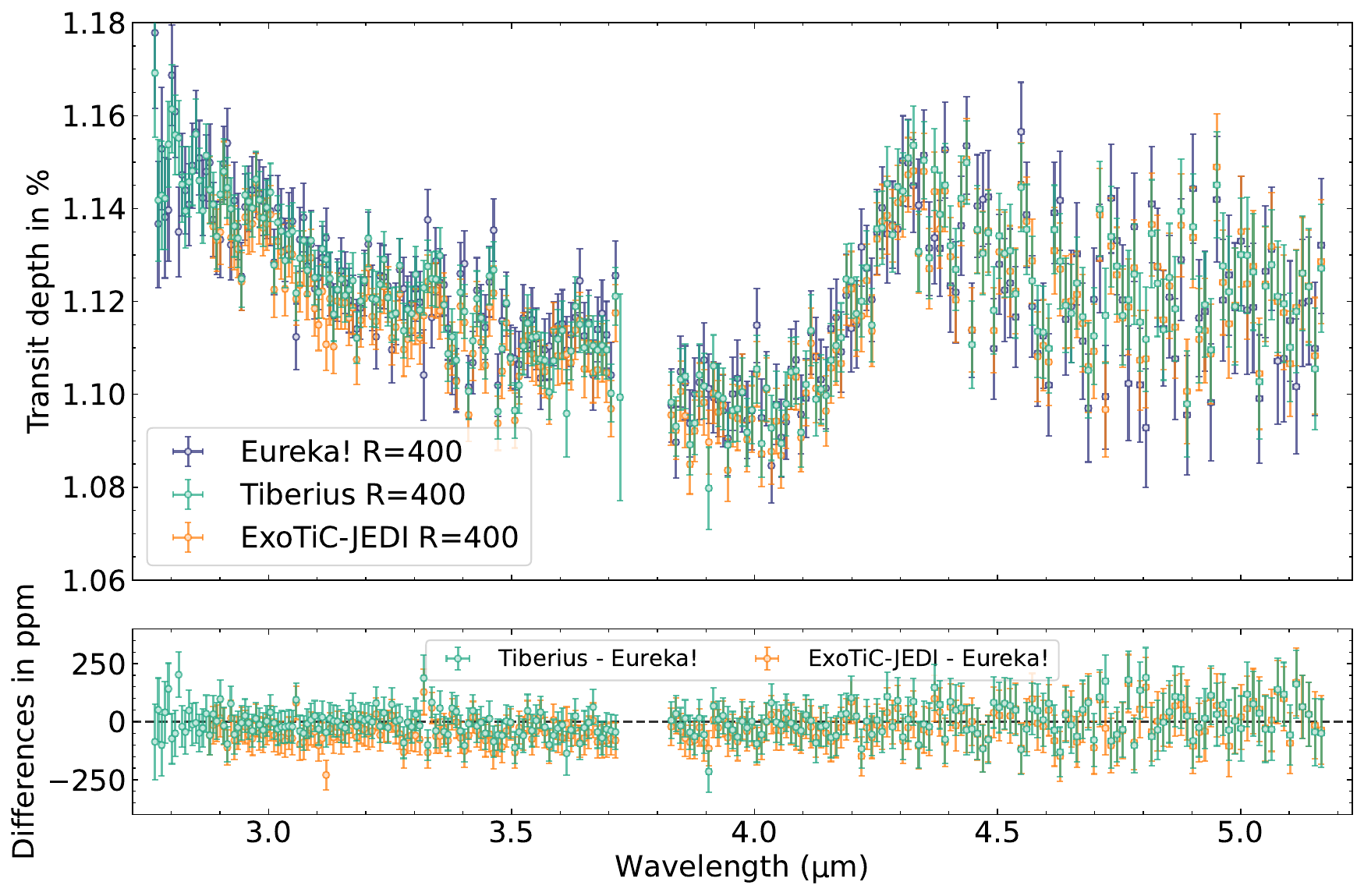}
    \end{minipage}
    \caption{\textit{Top panels: }Transmission spectrum of \myplanet using JWST's NIRSpec/G395H, using three independent reduction pipelines, \tiberius (green), \eureka (dark blue) and \exoticjedi (orange). The two panels show the spectra at two different binning schemes with $R=100$ on the left and $R=400$ on the right. \textit{Bottom panels:} Differences between the reductions compared to \texttt{Eureka!}. Note that the y axis on the left panel is a different scale (-150,+150)\,ppm compared to the right panel (-400,+400)\,ppm for visual clarity. \exoticjedi displays a slight offset compared to \tiberius and \eureka in the NRS1 spectrum.  }
    \label{fig:transmission-spectrum}
\end{figure*}

The spectroscopic light curves were fitted in the same manner as the white light curves, however, we held $a/R_*$, $i$ and $T_0$ fixed to the corresponding white light curve value for each detector as appropriate (see Table \ref{tab:system_params}). We used the same $R=100$ and $R=400$ binning schemes as the \eureka and \tiberius reductions, as shown in Figure \ref{fig:transmission-spectrum}. Note that the \exoticjedi reduction does not include bluest wavelength bins in NRS1 (see Fig.\,\ref{fig:transmission-spectrum}) as the throughput curve used in the computation of the limb-darkening values with \texttt{ExoTiC-LD} do not cover wavelengths $< 2.814$\,\microns, while \eureka and \tiberius reductions use an extrapolated throughput towards the bluer wavelength ranges. The official throughput files for G395H as provided by STScI G395H do not cover those bluer wavelengths and any information outside of the computed throughput of the instrument cannot be accurately assigned to any single wavelength and is potentially a combination of spectral wavelengths exposed on those pixels of the detector. So not using wavelength $< 2.814$\,\microns is a more conservative approach. 


\label{sec:trans_spec}

\subsection{Transmission spectra and limb asymmetries}

\begin{figure}
    \centering
    \includegraphics[width=\linewidth]{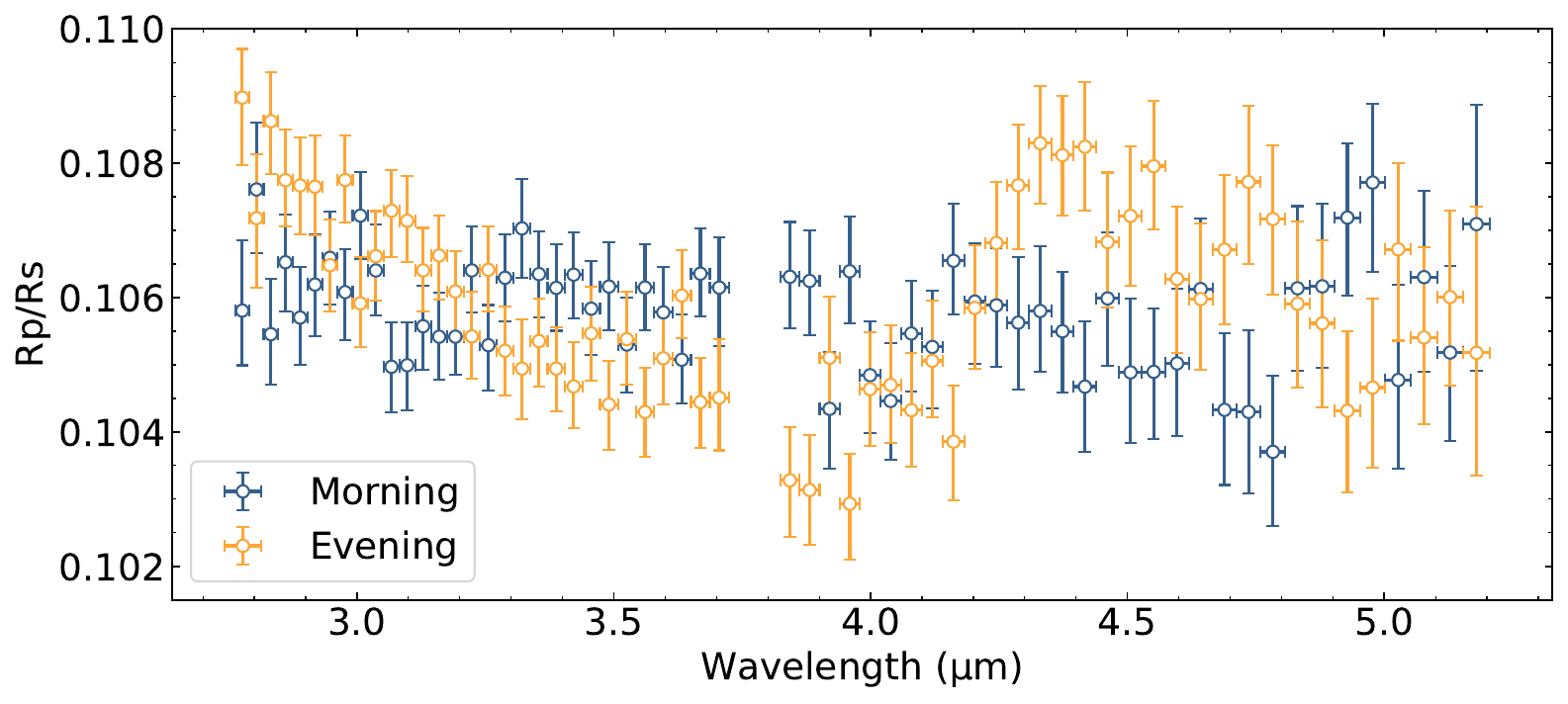}
    \caption{Morning \revision{(blue)} and evening \revision{(orange)} transmission spectrum of \myplanet applying \texttt{catwoman} \citep{Jones2021Catwoman:Curves,Espinoza2021ConstrainingSpectroscopy} to the \eureka R=100 light curves. }
    \label{fig:morning-evening}
\end{figure}

The resulting transmission spectra using NIRSpec/G395H for \myplanet from all three reductions, \eureka, \tiberius and \exoticjedi,  are shown in Fig.\,\ref{fig:transmission-spectrum}, at two spectral resolutions with $R=100$ and $R=400$ on the left and right panel, respectively, as well as their differences in the bottom panel. The data show visible absorption features of \ce{H2O} with the slope in the bluer end of the spectrum and \ce{CO2} at $\sim 4.4$\microns.

We additionally explore the possibility of limb asymmetries in our light curves using the \eureka R=100 light curves with the open-source python package \texttt{catwoman} \citep{Jones2021Catwoman:Curves,Espinoza2021ConstrainingSpectroscopy} and the nested sampling algorithm \texttt{dynesty} \citep{Speagle2020Dynesty:Evidences}. Asymmetries at the terminator region can arise from temperature differences and variations of chemical compositions on the morning and evening terminators of exoplanets, the presence of which has recently been demonstrated for hot gas giants using JWST \citep{Espinoza2024InhomogeneousB,Murphy2024EvidenceWASP-107b}. 

With \texttt{catwoman} we modelled the transiting planet as two semicircles with different radii. We fixed the orbital parameters for NRS1 and NRS2 as described in Table\,\ref{tab:system_params} and fixed the quadratic limb-darkening coefficients to the values from \texttt{ExoTIC-LD} \citep{Wakeford2022Exo-TiC/ExoTiC-LD:Release,Grant2024ExoTiC-LD:Coefficients} as described before. We fit the transit mid-time for each bin \revision{(Gaussian prior, from Table\,\ref{tab:system_params})} as well as the transit depths \revision{components as $R_\mathrm{p}/R_\mathrm{s}$ for morning and evening side (Uniform prior, $0.10 - 0.11$)} and an error inflation term. Before fitting with \texttt{catwoman} we remove the polynomial trend term found with the traditional light curve fitting. Therefore each spectroscopic bin is fitted using 4 free parameters: the morning and evening radius of the planet, mid-transit time and the noise term. We run the nested sampling algorithm with a total of 400 live points (100$\times$ number of dimensions). We further fit the same model without asymmetries, i.e., assuming the morning and evening radius to be equal (so 3 free parameters), to allow for Bayesian evidence comparison. 

The resulting morning and evening transmission spectra applying \texttt{catwoman} to the \eureka R=100 light curves are shown in Fig.\,\ref{fig:morning-evening}. We find different spectra between the two limbs, however, when comparing the Bayesian evidence difference between the asymmetric morning-evening model and the conventional symmetric terminator model, the latter, less complex model is sufficient to fit the data \citep[e.g.,][]{Jeffreys1983TheoryProbability,Trotta2008BayesCosmology}. In fact, the Bayesian evidence favours the symmetric transit model in all spectroscopic bins, with differences $\Delta \ln \mathcal{Z}$ ranging from $0.5-1.5$. \revision{The shape of the morning and evening spectra are consistent across reductions (see Appendix\,\ref{sec:appendix_catwoman-all-reductions}). In a similar manner, the small differences in Bayesian evidence $\ln \mathcal{Z}$ hold true for \tiberius and \exoticjedi as well, favouring the simpler model.} Future work may compare and contrast these potential limb asymmetries, but this is beyond the scope of this presented work.

\section{Atmospheric retrieval analysis }
\label{sec:retrievals}

We perform atmospheric retrievals using two independent retrieval codes -- \pRT and \hydra -- on our JWST transmission spectra of \myplanet. We use \pRT to run equilibrium chemistry retrievals as well as free chemistry retrievals with an isothermal temperature profile. We then employ the \hydra retrieval setup to investigate a non-isothermal profile and a more complex haze and cloud parameterisation. We further test for an offset between the two detectors, NRS1 and NRS2. 

\subsection{\pRT: free chemistry and equilibrium chemistry}
We performed free chemistry and equilibrium chemistry retrievals with \texttt{petitRADTRANS} (\pRT) version 3 \citep{Molliere2019PetitRADTRANS:Retrieval,Nasedkin2024AtmosphericPetitRADTRANS} on \myplanet's transmission spectra, both for R=100 and R=400 of the JWST spectra for all three reductions. \pRT explores the parameter space using the \texttt{Python} version of the nested-sampling algorithm \texttt{MultiNest} \citep{Feroz2009MULTINEST:Physics}. 

We included the following species using correlated-$k$ radiative transfer with opacity tables at R=1,000:  \ch{H2O}, \ch{CO}, \ch{CO2},  \ch{CH4}, \ch{H2S}, \ch{HCN}, \ch{NH3} and \ce{C2H2}. The individual references for each species can be found in Appendix, Table\,\ref{tab:ret_priors}. We assume an \ch{H2}- and He-dominated atmosphere and include opacity from \ch{H2-H2} and \ch{H2-He} collision-induced absorption \citep{Richard2012NewCIA}. \revision{The atmospheric pressures we consider range from $10^{-8}$ bar to $10^{2}$ bar. }

For the cloud parameterisation, we include a grey cloud deck with the cloud-top pressure as a free parameter. 
We fit for reference pressure and planet mass using a Gaussian prior centred on the planet's mass ($0.456$\,M$_\textrm{Jup}$) with a standard deviation of $0.05$\,M$_\textrm{Jup}$, while we fix the planet radius to $0.106\,\times$ the stellar radius (based on the white-light curve fit). We use a wide, uniform prior for the limb temperature. The stellar radius was fixed to 1.5784\,$R_{\odot}$ (see Table \ref{tab:wasp-94_parameters}). 
In the case of our free chemistry retrievals, we used priors for the individual log mass fractions of each species from -10 to 0.
In the equilibrium chemistry case, we used a uniform prior for the C/O ratio of $0.1-1.5$ (varying the oxygen content) and a log-uniform prior of -2 to 3 for the metallicity (Fe/H $\times$ solar). For the runs where an offset between NRS1 and NRS2 was fit, we employed a uniform prior from -200~ppm to +200~ppm. 
The priors of all parameters are summarised in the Appendix, Table\,\ref{tab:ret_priors}. In total, including the offset we fit 13 free parameters in the free chemistry and 7 parameters in the equilibrium chemistry case.

\subsection{\texttt{HyDRA}: free chemistry}

We perform free-chemistry atmospheric retrievals with \hydra \citep{Gandhi2019HyDRA-H:Spectra, Gandhi2022SpatiallyPhase}. This assumes the volume mixing ratio (VMR) of each species is a free parameter. Our molecular opacities are calculated from pre-computed grids over a range of pressures and temperatures spanning the photosphere of \myplanet. We include the same species as that for the \pRT retrievals (\ch{H2O}, \ch{CO}, \ch{CO2},  \ch{CH4}, \ch{H2S}, \ch{HCN}, \ch{NH3} and \ce{C2H2}). The individual line lists we used are summarised along the priors in Appendix, Table\,\ref{tab:ret_priors}. In addition to sources of line opacity, we also include absorption from H$_2$-H$_2$ and H$_2$-He interactions \citep{Richard2012NewCIA}. 

The \hydra retrieval includes six free parameters which determine the vertical thermal profile of the atmosphere, following the procedure by \citet{Madhusudhan2009AAtmospheres}. \revision{We divide the atmosphere into 50 equal layers in log-space with pressures from $10^{-7}$ bar to $10^{2}$ bar.} This allows for a range of non-inverted, inverted and isothermal profiles driven by the observational constraints. We also include an additional parameter for the reference pressure at which the radius of \myplanet is set. For cloud and haze contributions to the opacity we include four additional parameters, including the cloud fraction, as discussed in \citet{Welbanks2021Aurora:Spectra} and \citet{Gandhi2022SpatiallyPhase}. Finally, we retrieve an offset between the two NIRSpec detectors, resulting in a total of 20 free parameters in the retrieval. Appendix Table~\ref{tab:ret_priors} shows the prior ranges for each of the parameters. Our retrievals are performed using the \texttt{MultiNest} Nested Sampling algorithm \citep{Feroz2008MultimodalAnalyses, Feroz2013BayesianPlanets, Buchner2014X-rayCatalogue}, with the spectral model generated at R=100,000 and then convolved and binned to the data resolution.

\begin{figure*}
    \centering
    \includegraphics[width=\linewidth]{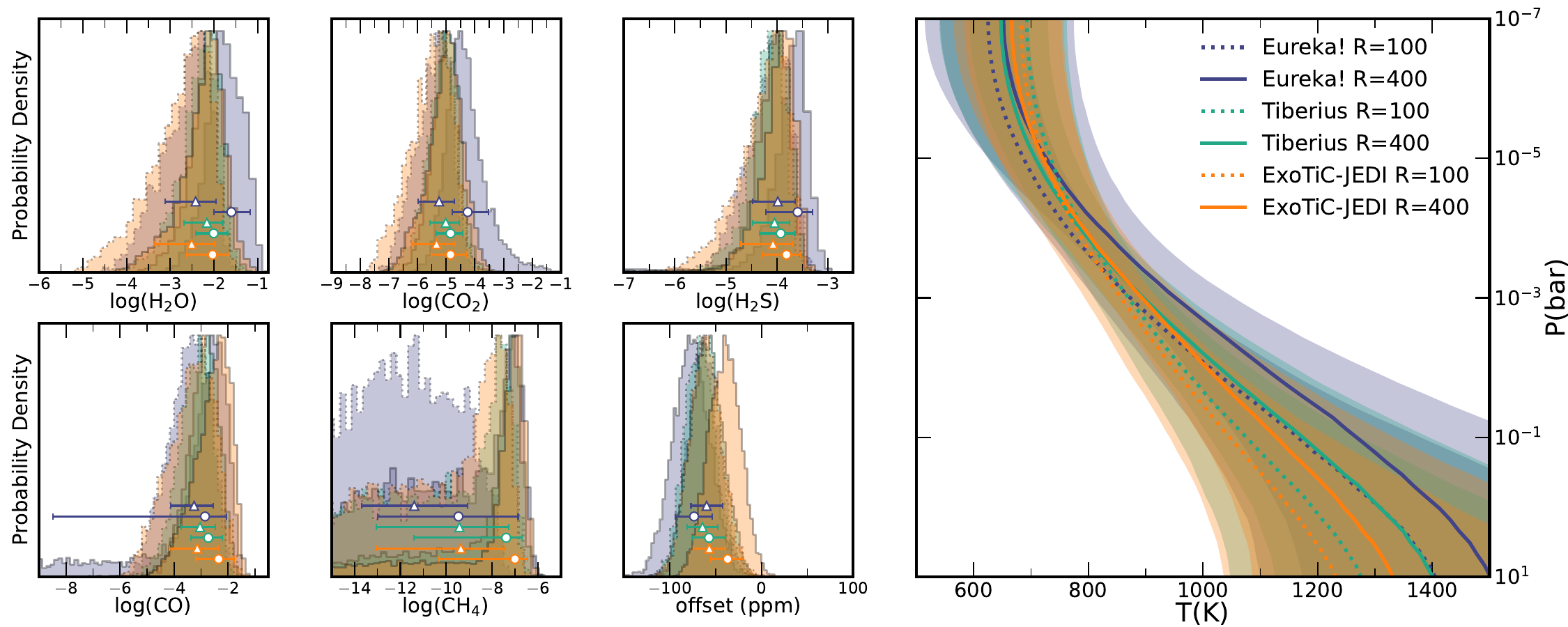}
    \caption{Constrained retrieval parameters for \myplanet from \hydra for each of the reductions at both binned spectral resolutions. In the left panels we show the volume mixing ratios for the chemical species and the offset (in ppm) between the two detectors and in the right panel we show the constrained temperature profile, with the median and the $\pm 1\sigma$ range.}
    \label{fig:ret}
\end{figure*}

\begin{figure}
    \centering
    \includegraphics[width = \linewidth]{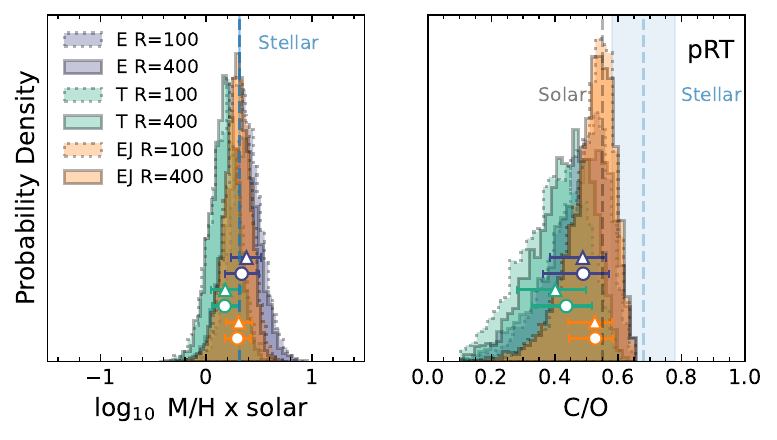}
    \caption{Retrieved metallicity and C/O constraints for \myplanet from the equilibrium chemistry retrievals using \pRT for each of the reductions at both binned spectral resolutions. The round circles and triangle shapes as markers correspond to the median value that was found for the R=100 and R=400 resolutions, respectively. }
    \label{fig:ret_co_met_eq_chem}
\end{figure}

\begin{figure}
    \centering
    \includegraphics[width=\linewidth]{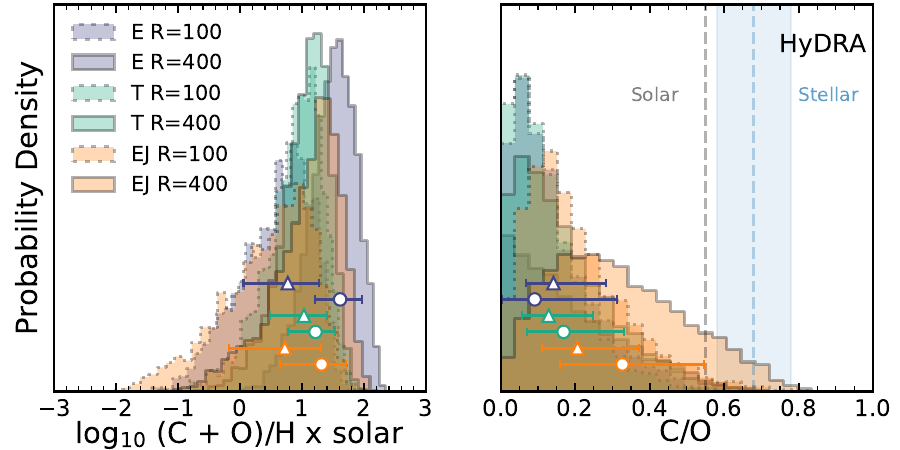}

    \includegraphics[width=\linewidth]{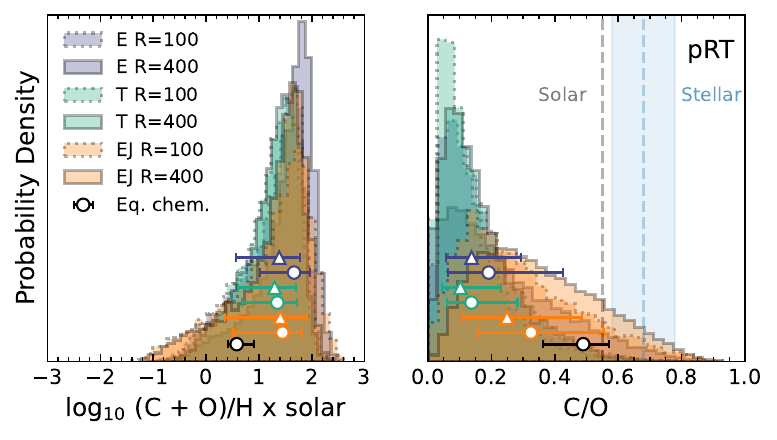}
    
    \caption{Derived metallicity as (C+O)/H compared to the solar (C+O)/H, and C/O constraints for \myplanet from the two free chemistry retrievals; \hydra (top) and \pRT (bottom) for each of the reductions at both binned spectral resolutions. On the bottom panels we also include the respective equilibrium chemistry values derived by \pRT for \eureka R=400. }
    \label{fig:ret_co_met}
\end{figure}

\begin{figure*}
    \includegraphics[width=\linewidth]{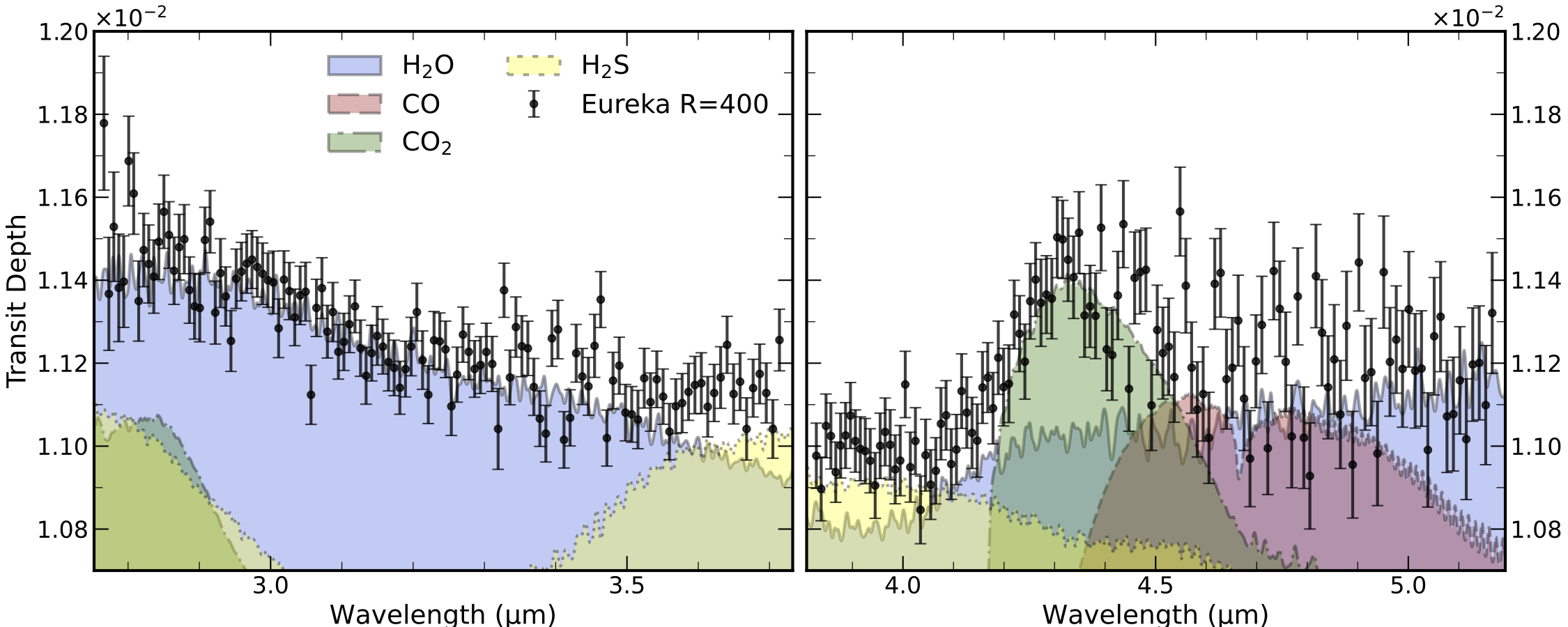}
    \caption{Transmission spectrum of \myplanet using JWST's NIRSpec/G395H with the \eureka R=400 dataset in black and the best \hydra fitting retrieval model with each species modelled individually. We show the contribution to the spectrum for the four species that showed a positive detection significance, namely H$_2$O, CO, CO$_2$ and H$_2$S.}
    \label{fig:ret_contribution}
\end{figure*}












\section{Results and Discussion}
\label{sec:discussion}

\subsection{Atmospheric characterisation of \myplanet}

For \myplanet's atmosphere, we retrieve consistent parameters for mass and reference pressure across both chemical equilibrium and free chemistry retrievals and between the two retrieval codes. 

Temperatures and cloud-top pressures varied between the chemical setups in \pRT and \hydra. In both the free chemistry retrievals the temperatures are lower ($\approx 200\,K$ lower) and the cloud deck is at a higher pressure ($\Delta \log P_\mathrm{cloud} \sim 2$ bar, i.e., higher up in the atmosphere) compared to the equilibrium setup. These two parameters are degenerate, as to first order the scale height (and thus the features) increase with temperature while the cloud-top pressure mutes the features when it decreases. 

In all the equilibrium chemistry retrievals we find a limb temperature of $900 - 1000$\,K across all reductions (with uncertainties of $\approx 100-200$\,K), which is lower than the equilibrium temperature. This is a common occurrence in 1D isothermal retrievals due to the unresolved limb asymmetries \citep{MacDonald2020WhyAtmospheres} and/or from the PT profile parametrisation \citep{Welbanks2022OnTerminators}.

\subsubsection{Evidence for an offset between NRS1 and NRS2}
\label{sec:retrieval-offset-implications}
While NRS1 and NRS2 are the same type of detectors, they do exhibit performance differences. In addition, NRS1 is more often illuminated compared to NRS2 as NRS2 is not used for every observing mode while NRS1 is. This results in differences in the systematics properties, e.g., as seen in time-series observations where the NRS1 light curves show a stronger linear trend compared to the NRS2 equivalent. This is also the case for our data set, though with the deep transit of \myplanet it is hardly visible in Fig.\,\ref{fig:WL_curves}. Due to this effect, it has become common practice to test for detector offsets in the transmission spectrum by investigating the inclusion of an offset parameter in the atmospheric retrievals \citep[e.g.,][]{May2023DoubleG395H, Ohno2024ASpectrum,Alderson2025JWSTTOI-776b,Bello-Arufe2025EvidenceL98-59b,Schmidt2025ASpectrum}. 

With both \pRT and \hydra we run retrievals for equilibrium and free chemistry including and excluding an offset between NRS1 and NRS2 for all reductions at all resolutions. We find that in all cases the Bayesian evidence comparison favours the model with the offset over one without an offset, see Table\,\ref{tab:bayesian_evidences}. The values of the offset fitted by all retrievals agree well for all reductions and are of the order of $38-93$\,ppm, see Table\,\ref{tab:offsets}. 

The statistical significances of preference for an offset vary slightly across reductions and retrieval set-ups. The Bayesian evidence difference $\Delta\ln\mathcal{Z}$ range from $ 2.3 - 7.2$ ($2.95\sigma$ -- $4.21\sigma$) for the \eureka and \tiberius reductions, clearly favouring the retrievals with the offsets. The fitted offsets using the \eureka and \tiberius reductions also agree within their $1\sigma$ uncertainties. 
On the other hand, the retrievals on the \exoticjedi reductions do not show a clear preference, with model evidence differences of $ 0.3 - 1.5$ ($<1\sigma$), though still in favour of including offsets. In addition, all fitted offsets are inconsistent with a zero offset at a $2.1-3.5\sigma$ level.


We note that the C/O ratio and abundances in the chemical equilibrium and free chemistry, respectively, are affected by the choice of whether an offset is included or not. For example, the C/O ratio for our fiducial spectrum (\eureka, R=400) without an offset is C/O $=0.194 \pm 0.079$, whereas if we include an offset in the retrieval the C/O $= 0.49^{+0.08}_{-0.13}$. 
We also find evidence for \ce{CH4} at a $4\sigma$ level which seems unlikely for a planet at this temperature. Therefore we urge the community to test detector offsets when running atmospheric retrievals on JWST NIRSpec/G395H observations. 

In summary, we see a slight offset between the data reductions by eye in Fig.\,\ref{fig:transmission-spectrum}, the Bayesian evidence prefers the inclusion of an offset for all reductions at both resolutions and for all retrieval setups, and the fitted offsets are inconsistent with a zero at a $2.1-3.5\sigma$ level. We conclude that the inclusion of an offset between NRS1 and NRS2 is necessary and therefore all our fiducial models include an offset between NRS1 and NRS2.

\begin{table*}
    \centering
    \caption{Bayesian evidences differences $\Delta \ln\mathcal{Z}$ for \revision{equlibrium and free chemistry} retrieval setups with and without an offset between the two NIRSpec NRS1 and NRS2 detectors. The \pRT values \revision{(both equilibrium and free chemistry)} are relative to the favoured retrieval, which is the equilibrium chemistry setup including the offset (indicated by '--'\revision{, first column}). \revision{Therefore this also demonstrates that the equilibrium chemistry setup with offset is preferred over the free chemistry.} The \revision{two} rightmost columns correspond to the $\Delta \ln\mathcal{Z}$ values for the \hydra free chemistry setup, \revision{relative to the setup with an offset ('--'),} also showing a preference for including an offset. }
    \label{tab:bayesian_evidences}
    \begin{tabular}{lccccccc}
    \hline
           &  \multicolumn{2}{c}{\pRT: Equilibrium Chemistry}  &  \multicolumn{2}{c}{\pRT: Free Chemistry} & &  \multicolumn{2}{c}{\hydra: Free Chemistry} \\ \hline
        Reduction & w/ offset & w/o offset & w/ offset & w/o offset & &  w/ offset & w/o offset \\ \hline
        \eureka R=100 & -- & $-5.2\pm 0.2$ &  $-2.7\pm 0.2$ & $-7.3\pm 0.2$ & & -- & $-3.2 \pm 0.2 $\\
        \eureka R=400 & -- & $-7.2\pm 0.2$ & $-3.7\pm 0.2$ & $-7.9\pm 0.2$ & & -- & $-3.7 \pm 0.2 $\\
        \tiberius R=100 & -- & $-2.3\pm 0.2$ &  $-1.6\pm 0.2$ & $-5.1\pm 0.2$ & & -- & $-3.4 \pm 0.2 $\\
        \tiberius R=400 & -- & $-2.9\pm 0.2$ & $-2.8\pm 0.2$  & $-5.8\pm 0.2$ & & -- & $-2.9 \pm 0.2 $\\
        \exoticjedi R=100 & -- & $-1.5\pm 0.2$ &  $-3.0\pm 0.2$ & $-5.2\pm 0.2$ & & -- & $-2.6 \pm 0.2 $\\
         \exoticjedi R=400 & -- & $-0.4\pm 0.2$ &  $-4.4\pm 0.2$ & $-4.7\pm 0.2$ & & -- & $-0.6 \pm 0.2 $\\ \hline 
    \end{tabular}

\end{table*}

\begin{table}
    \centering
    \caption{Retrieved offsets between the two NIRSpec/G395H detectors, NRS1 and NRS2. In \pRT it is applied as a subtraction to the transit depths in NRS2 so the negative value indicates that NRS2 is shifted up in regards to NRS1. The full retrieval results for these runs are shown in Table\,\ref{tab:full-retrieval-results}. }
    \label{tab:offsets}
    \begin{tabular}{lccc}
    \hline
    \multicolumn{4}{c}{Retrieved offset between NRS1 and NRS2 in ppm} \\ \hline
    Reduction     &  \pRT eq.~chem. & \pRT free~chem. & \hydra \\ \hline
     \eureka R=100    & $-69 \pm 15$  & $-76 \pm 18$ & $-60 \pm 17$  \\
     \eureka R=400   & $-87 \pm 17$  & $-93 \pm 21$ & $-74 \pm 20$  \\
     \tiberius R=100    & $-60 \pm 17$  & $-80 \pm 19$ & $-64 \pm 17$   \\ 
    \tiberius R=400    & $-57 \pm 16$  & $-76 \pm 20$ &$-57 \pm 18$  \\
     \exoticjedi R=100    & $-52 \pm 16$  & $-69 \pm 20$  &  $-57 \pm 17$ \\
     \exoticjedi R=400    & $-47 \pm 16$  & $-52 \pm 22$ & $-37 \pm 18$  \\ \hline
    \end{tabular}
    
\end{table}

\subsubsection{C/O and metallicity}

\begin{table*}
\def\arraystretch{1.2}
\caption{Retrieved C/O ratios and metallicities of \myplanet's atmosphere in the chemical equilibrium case (\revision{left}) and the inferred values from the free chemistry retrievals (\revision{right}), demonstrating that the equilibrium chemistry prefers higher C/O and lower metallicity. Solar refers to \citet{Asplund2009TheSun}. The (C+O)/H ($\times$solar) for all \pRT equilibrium runs were computed using the in-built conversion to O/H and C/H, where the value stated here is using the median C/O and median metallicity for each reduction and the uncertainties stated here use both $1\sigma$ C/O and metallicity uncertainties to determine a conservative boundary (i.e., the upper error is derived using the highest (C+O)/H, with the $+1\sigma$ of metallicity (varying C) and the $-1\sigma$ of C/O (varying O) and vice versa for the lower error). Note that Bayesian evidence favours the equilibrium chemistry models for all reductions. }
    \label{tab:CtoO_mets}
    \centering
    \begin{tabular}{lccccccc}
    \hline
     & \multicolumn{3}{c}{\textbf{Equilibrium chemistry:} \pRT}     &   \multicolumn{2}{c}{\textbf{Free chemistry:} \pRT} &  \multicolumn{2}{c}{\textbf{Free chemistry:} \hydra} \\  \hline
     Reduction & C/O  & Z ($\times$solar) & (C+O)/H ($\times$solar) & C/O & (C+O)/H ($\times$solar) & C/O & (C+O)/H ($\times$solar) \\ \hline
     \eureka R=100    & $0.49^{+0.07}_{-0.10}$  & $ 2.41^{+0.96}_{-0.65} $ & $4.2^{+2.6}_{-1.4}$& $0.13^{+0.16}_{-0.09}$  & $24^{+36}_{-21}$ & $0.14^{+0.14}_{-0.07}$ &  $6^{+13}_{-5}$\\
     \eureka R=400   & $0.49^{+0.08}_{-0.13}$ & $2.17^{+0.96}_{-0.68}$ & $3.8^{+2.8}_{-1.4}$ & $0.19^{+0.24}_{-0.13}$ & $46^{+48}_{-36}$ &  $0.09^{+0.22}_{-0.09}$ &  $41^{+50}_{-25}$\\
     \tiberius R=100    & $0.40^{+0.10}_{-0.12}$  & $1.52^{+0.54}_{-0.44}$ & $3.0^{+2.2}_{-1.2}$ &  $0.10^{+0.13}_{-0.06}$ & $20^{+31}_{-16}$ &  $0.13^{+0.12}_{-0.07}$ &  $11^{+14}_{-8}$ \\
    \tiberius R=400     & $0.44^{+0.08}_{-0.11}$ & $1.50^{+0.48}_{-0.41}$& $2.79^{+1.7}_{-0.95}$ & $0.14^{+0.15}_{-0.08}$ & $22^{+31}_{-18}$ &  $0.17^{+0.16}_{-0.10}$ &  $16^{+18}_{-11}$ \\
     \exoticjedi R=100    & $0.526^{+0.054}_{-0.083}$ & $2.04^{+0.65}_{-0.50}$ & $3.38^{+1.6}_{-0.96}$ & $0.25^{+0.24}_{-0.14}$ & $25^{+56}_{-24}$ &  $0.21^{+0.17}_{-0.10}$ &  $5^{+15}_{-5}$  \\
     \exoticjedi R=400    & $0.527^{+0.056}_{-0.084}$ & $1.97^{+0.59}_{-0.49}$ & $3.28^{+1.4}_{-0.96}$ & $0.32^{+0.23}_{-0.17}$ & $28^{+38}_{-25}$ & $0.33^{+0.22}_{-0.17}$ &  $20^{+30}_{-16}$\\ \hline
    \end{tabular}
\end{table*}

With our equilibrium chemistry retrieval analysis using \pRT, we find the ratio between carbon and oxygen molecules in the atmosphere of \myplanet to be subsolar to solar (ranging from $0.40 - 0.53$, see Table\,\ref{tab:CtoO_mets}) and consistent across reductions within their $1\sigma$ uncertainties, see Fig.\,\ref{fig:ret_co_met_eq_chem}. 
For the atmospheric metallicity of \myplanet we find slightly supersolar metallicities for all reductions and resolutions, ranging between 1.5 and 2.4 $\times$ solar.

\begin{figure}
    \centering
    \includegraphics[width=\linewidth]{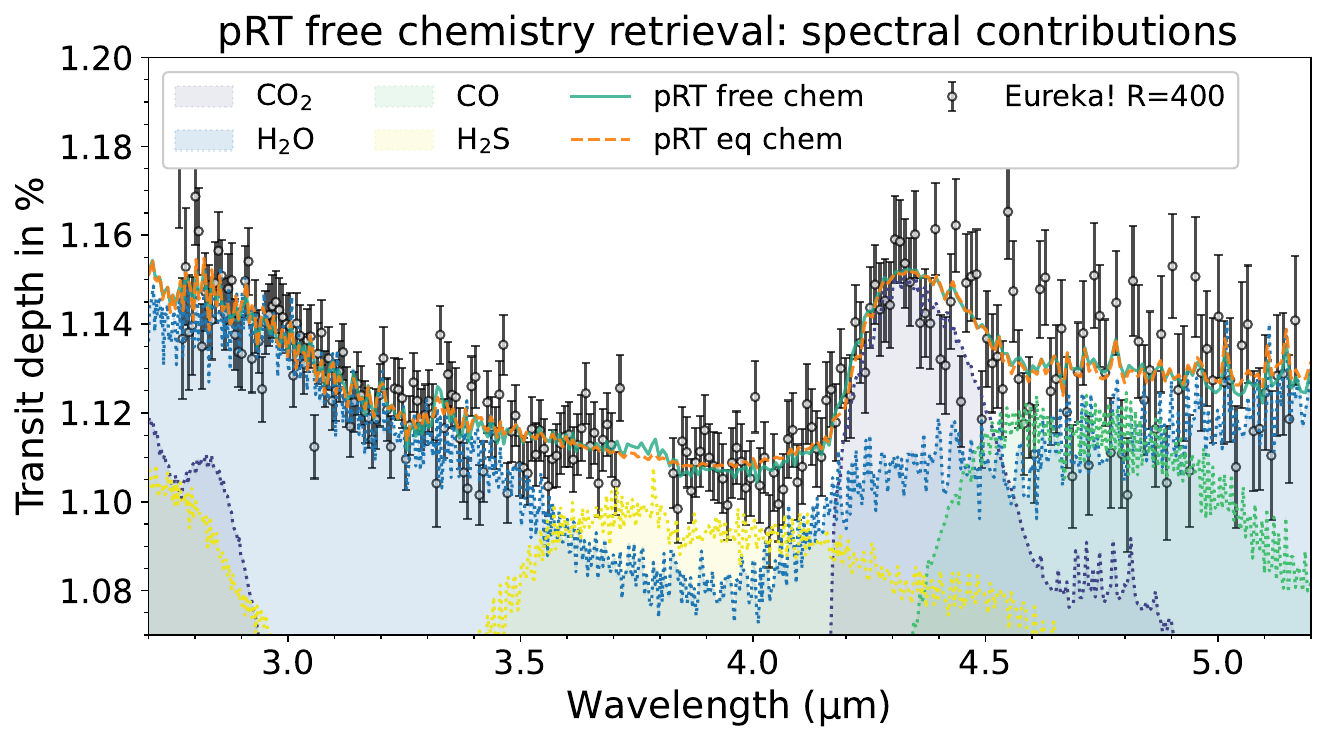}
    \includegraphics[width=\linewidth]{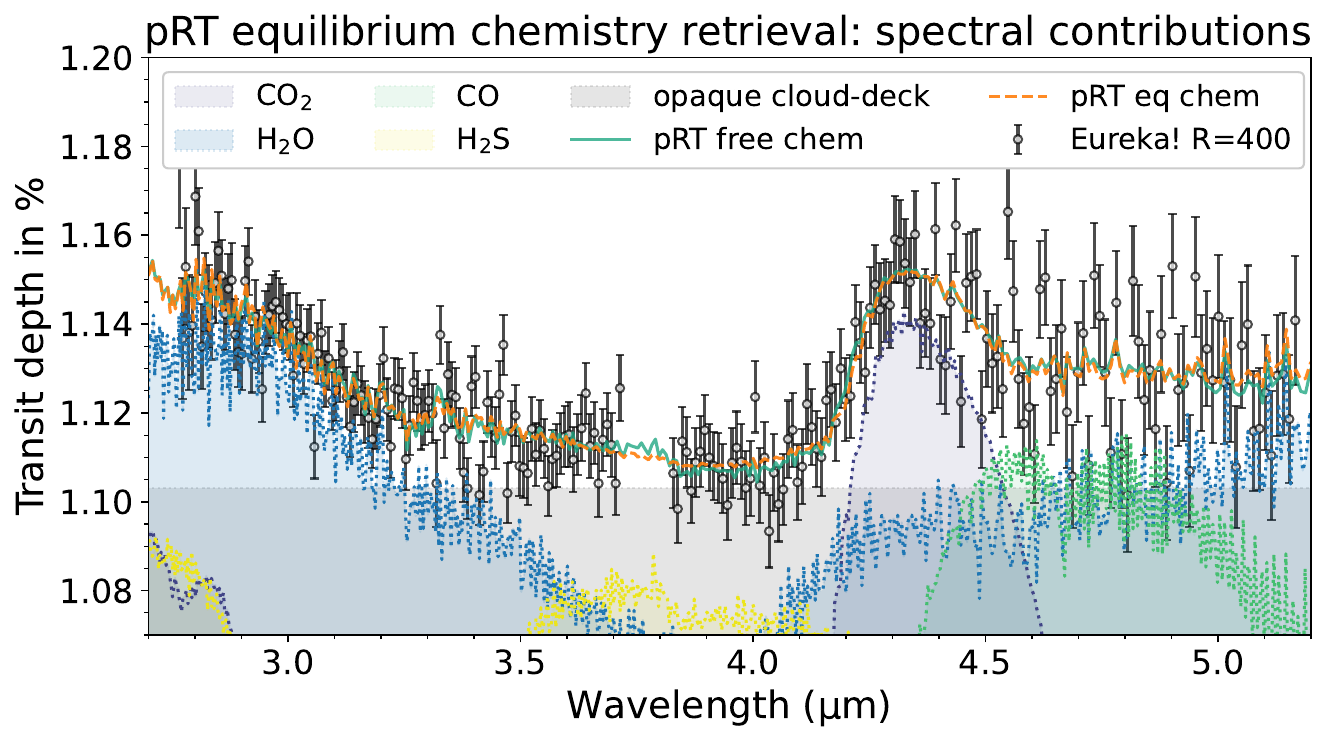}
    \caption{\eureka transmission spectrum of \myplanet with the best-fit \pRT models using free chemistry and equilibrium chemistry and the fitted corresponding spectral contributions on the top and bottom panels, respectively. The best-fit offset from the latter retrieval is applied to the NRS2 spectrum (87ppm, consistent within $0.3\sigma$ with the 92ppm from the free chemistry run). By eye the two models (equilibrium chemistry vs free chemistry) show remarkable agreement with the exception of the cloud-deck which is at much higher altitudes for the equilibrium chemistry (see bottom panel), while the free chemistry retrieval prefers a cloud deck at lower altitudes not shown in the top panel as it is off the y axis (equivalent of $\sim 1.05 \%$ transit depth). However, the equilibrium chemistry retrieval runs were preferred across the board when comparing their Bayesian evidence values for all reductions and resolutions, see Table\,\ref{tab:bayesian_evidences}. Therefore, our favoured model is the equilibrium model, with $\Delta \ln \mathcal{Z} = 3.7$ ($3.2\sigma$) in the case of \eureka R=400 using \pRT.  }
    \label{fig:spectrum-free_chem-vs-eq_chem}
\end{figure}

For comparison, we computed the C/O ratio and metallicities for our best-fit free chemistry models for all reductions and resolutions using \hydra and \pRT, summarised in Table\,\ref{tab:CtoO_mets} and shown in Fig.\,\ref{fig:ret_co_met}. The best-fit using \hydra is shown in Fig.\,\ref{fig:ret_contribution}, and the two best-fit free chemistry and equilibrium chemistry models using \pRT are shown in Fig.\,\ref{fig:spectrum-free_chem-vs-eq_chem}. The numbers computed by \hydra and \pRT are consistent within $<1\sigma$ for both the C/O ratios and metallicity (as (C+O)/H $\times$ solar).  

Across the board, we find that the C/O values in the \hydra and \pRT free chemistry retrievals are lower than the ones retrieved by the \pRT chemical equilibrium retrievals, ranging from $0.10 - 0.32$ versus $0.40 - 0.53$, although the uncertainties on those computed values are relatively high and therefore consistent with the equilibrium chemistry model. In addition, using free chemistry abundances to compute C/O ratios may be biased towards lower C/O as features by carbon species such as \ce{CH4} as well as \ce{HCN} (and \ce{CO}) are blended with the strong \ce{H2O} bands and are not fully spectrally resolved. Therefore their abundances may be underestimated in free chemistry retrievals and contributing to a lower C/O ratio.   

\begin{figure}
    \centering
    \includegraphics[width=0.9\linewidth]{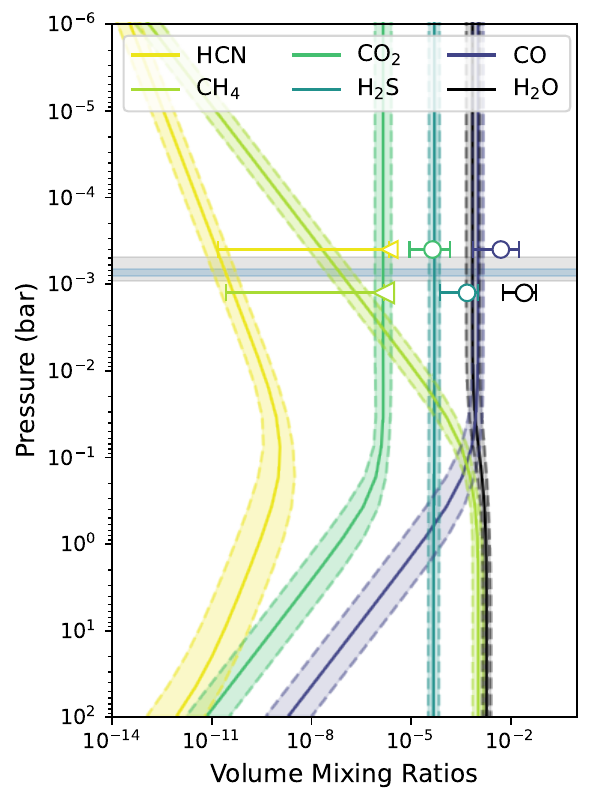}
    
    \caption{Abundances of the molecules versus pressure in \myplanet's atmosphere as computed by the best-fit equilibrium chemistry (coloured lines) overlayed with the best-fit abundances (circles/triangles, $2\sigma$ \revision{ranges as} upper limits for \ce{CH4} and \ce{HCN}) as determined by the free chemistry model (at arbitrary pressures), both computed with \pRT, see also Fig.\,\ref{fig:spectrum-free_chem-vs-eq_chem}. The reference pressures of the two models are also indicated by the gray (equilibrium chemistry) and blue (free chemistry) shaded box.  }
    \label{fig:vmrs-vs-pressure}
\end{figure}

In contrast, the inferred metallicities are higher compared to the retrieved values by the equilibrium chemistry run. When comparing the volume mixing ratios (see Fig.\,\ref{fig:vmrs-vs-pressure}), it is clear that the free chemistry retrievals show consistently higher abundances for all molecules compared to the equilibrium chemistry model. Therefore the (C+O)/H will be higher for the free chemistry. The difference in abundances can be explained by the difference in cloud deck (see Table\,\ref{tab:full-retrieval-results}), where the equilibrium chemistry retrieval prefers the grey cloud deck to be higher up in the atmosphere and therefore muting the features. 

Using the Bayesian evidence \revision{values for all atmospheric retrieval runs, we find that the equilibrium chemistry models are preferred across all reductions and all resolutions. This is visible in Table\,\ref{tab:bayesian_evidences}, where the \pRT equilibrium chemistry runs are preferred over the more complex free chemistry runs with $\Delta \ln\mathcal{Z} 1.6 - 4.4$.} Thus, we do not detect evidence that points to disequilibrium processes \revision{in \myplanet's atmosphere. This is not surprising as there are only minimal differences visible when inspecting the best-fit equilibrium and free chemistry models in Fig.\,\ref{fig:spectrum-free_chem-vs-eq_chem}, and equilibrium chemistry requires} a smaller number of free parameters.

In summary, for \myplanet's atmosphere we find subsolar to solar C/O ratios, with the chemical equilibrium values being higher. The inferred atmospheric metallicities are consistent with solar at $1-2 \sigma$ levels though preferred to be higher, with a range of $1.5-2.4 \pm 1.0 \times$ solar when using the equilibrium chemistry results of all reductions at both resolutions. The retrieved C/O ratio and metallicity of our fiducial spectrum (\eureka, R=400) using our preferred chemical equilibrium model is C/O $= 0.49^{+0.08}_{-0.13}$ and Z $=2.17 \pm 0.65\ \times$ solar.   

As has been discussed in other works \citep{Welbanks2019MassMetallicityK,Reggiani2022}, it is advisable to compare planetary abundances/abundance ratios to those of their specific host star. \cite{Teske2016THEB} focused on the abundances of WASP-94\,A relative to WASP-94\,B, but here we are interested in just the C/O ratio of WASP-94\,A. To derive this value, we use (1) the absolute abundance ranges for oxygen and carbon quoted in \cite{Teske2016THEB} for WASP-94\,A from synthesis fitting blended lines (see their Section 3.2) as well as (2) previously unreported absolute abundance ranges measured from equivalent width fitting to unblended O I, C I, and CH lines (see Appendix\,\ref{sec:appendix_O_C_lines} for more details). Both of these approaches yield consistent results of a $(C/O)_*$ ratio of $\sim$0.68, using the solar values of carbon and oxygen abundances from \cite{Asplund2021}. Encompassing the different carbon and oxygen indicators as well as errors on the stellar parameters, we use a conservative error and proceed with $(C/O)_*$=0.68$\pm$0.10 for WASP-94\,A. This value is in line with what is predicted for a high metallicity star like WASP-94\,A \citep[e.g.,][]{Nissen2014}.

Together these show that the planet's C/O ratio is substellar, at $0.72 \pm 0.17 \times (C/O)_*$. \citet{Teske2016THEB} also measured a precise metallicity for \mystar of $2.09 \pm 0.02\ \times$ solar (0.32\,dex) which is in excellent agreement with our value for \myplanet's atmospheric metallicity of $2.17 \pm 0.67\ \times$ solar. Thus \myplanet's atmospheric metallicity is $1.04 \pm 0.33 \times (M/H)_*$.

\begin{table*}
\def\arraystretch{1.2}
\caption{Constrained volume mixing ratios for each molecular species from the best fit, where the uncertainties refer to the $1\sigma$ confidence intervals, while the upper limits refer to a $2\sigma$ interval. The detection significances for each detected species are shown as computed from the Bayesian evidence values based on the \eureka R=400 dataset retrievals with \hydra and \pRT.}
    \label{tab:hydra_prt_species_det_sig}
\begin{tabular}{lcccc}
    \hline 
     \eureka R=400 & \multicolumn{2}{c}{\hydra } & \multicolumn{2}{c}{\pRT }\\
    Species  &  log(Volume Mixing Ratio) & Detection significance ($\sigma$) & log(Volume Mixing Ratio) & Detection significance ($\sigma$)\\
\hline 
H$_2$O &  $-1.60^{+0.43}_{-0.40}$  &  4.0 & $-1.59^{+0.35}_{-0.64}$ & 4.1\\
CO &  $-2.85^{+0.79}_{-5.63}$  &  2.8 & $-2.30^{+0.55}_{-0.84}$ & 3.3\\
CO$_2$ &  $-4.25^{+0.74}_{-0.54}$  &  10.7 & $-4.35^{+0.52}_{-0.69}$ & 11.2\\
CH$_4$ &  $<-6.29$  &  -  & $<-5.83$ & -\\
NH$_3$ &  $<-5.10$  &  - & $<-5.07$ & -\\
HCN &  $<-5.67$  &  - &  $<-5.65$  & -\\
C$_2$H$_2$ &  $<-6.56$  &  - & $<-6.23$ & -\\
H$_2$S &  $-3.59^{+0.29}_{-0.62}$  &  2.9 & $-3.31^{+0.33}_{-0.80}$ & 2.1\\
\hline
    \end{tabular}

\end{table*}

\subsubsection{Molecular detections}

Using our free chemistry retrievals \hydra and \pRT, we investigate the absorbing molecules present in \myplanet's atmosphere. We ran retrievals excluding the individual species for our fiducial spectrum (\eureka, R=400) and computed the statistical significance via the Bayesian evidence model comparison. The detection significances are summarised in Table\,\ref{tab:hydra_prt_species_det_sig}. The two retrieval setups are consistent in their abundance constraints within less than $1\sigma$ and we find strong detection significance using \hydra and \pRT, respectively, for \ce{CO2} ($10.7\sigma$,$11.2\sigma$), good detection of \ce{H2O} ($4.0\sigma$,$4.1\sigma$) and tentative evidence for \ce{CO} ($2.8\sigma$,$3.3\sigma$) and \ce{H2S} ($2.9\sigma$,$2.1\sigma$).  

While we find very strong significance for \ce{CO2} for both retrievals, we only find $\sim4\sigma$ significance for \ce{H2O} for \hydra. This is in contrast with the high water abundance and therefore significant absorption detected, e.g., see Fig.\,\ref{fig:ret_contribution} and Table\,\ref{tab:hydra_prt_species_det_sig}. In the case of the \hydra retrieval, the low detection significance is explained by the fact that the model without \ce{H2O} is compensated by very high CO and CO$_2$ volume mixing ratios ($\log(\mathrm{VMR}_\mathrm{CO})\gtrsim-2.0$ and $\log(\mathrm{VMR}_\mathrm{CO_2})\gtrsim-3$ respectively), and an offset of $\sim200$~ppm. 
Similarly, \pRT finds high mass fractions for \ce{CO2}, \ce{NH3} and \ce{H2S}, unlikely to be physical and an offset of $\sim200$~ppm as well. If we fix the offset value to the value found by the base retrieval ($93$\,ppm, Table\,\ref{tab:offsets}) and redo our retrieval run with and without \ce{H2O} the detection significance increases to $11\sigma$ ($\Delta \ln \mathcal{Z} = 59$). Therefore, our quoted $4.1\sigma$ can be seen as a lower, conservative detection significance for \ce{H2O} driven by the uncertainty in the detector offset.  
The observations using JWST NIRISS/SOSS (GO 5924, PI: Sing) taken in October 2024 will be able to provide additional constraints on the water in the atmosphere of \myplanet and a combined study will be able to place further constraints on the C/O ratio and atmospheric metallicity. Combining the presented NIRSpec/G395H spectrum with the observations using NIRISS/SOSS may also help to constrain the offset as we get a better picture of the water abundance and clouds in the atmosphere.

We find tentative evidence for \ce{H2S} in the atmosphere of \myplanet. When we retrieve an atmosphere without \ce{H2S} with \pRT, it compensates for the absence of the molecule by pushing the cloud layer to lower pressures by $\sim$ two magnitudes and decreasing the water abundances, as well as increasing the offset between NRS1 and NRS2. Nevertheless, we find that the model with \ce{H2S} in the atmosphere is preferred by a $2.1\sigma$ and $2.9 \sigma$ significance by \pRT and \hydra, respectively. 
The \hydra retrievals without \ce{H2S} prefer very high CO and \ce{CO2} volume mixing ratios of $\log(VMR_\mathrm{CO})\gtrsim-1.5$ and $\log(VMR_\mathrm{CO_2})\gtrsim-3$, with an offset of $\sim200$~ppm. Hence, the model without \ce{H2s} is significantly less physically plausible than the model including \ce{H2S} which gives a metallicity of $\sim41\times$ metallicity for the \eureka R=400 dataset (see Table~\ref{tab:CtoO_mets}).
We find no evidence for the photochemical product \ce{SO2}, which is not expected given \myplanet's equilibrium chemistry and the relatively low metallicity we infer for its atmosphere.

\subsection{Implications for planet formation and migration}
\label{sec:implications-planet-formation-dynamics}

\subsubsection{Inferences from WASP-94's orbital dynamics}
The exact formation conditions of \myplanet are hard to deduce, although its highly oblique, retrograde orbit can provide some key clues about its dynamic history.
Firstly, the effective temperature of \mystar \citep[$6194 \pm 5$~K,][]{Teske2016THEB} places the star above the Kraft break \citep[around $\sim 6000$~K, dependent on stellar metallicity,][]{Kraft1967:RotationBreak}, which delineates the separation between hot, quickly rotating stars and cooler stars with thicker convective envelopes \citep[e.g.,][]{Dawson2014:HJTidal, Albrecht2022StellarSystems}. 
\myplanet is consistent with the wider observed trend of hot stars hosting hot Jupiters with higher obliquities than cooler stars.

Secondly, the combination of the high obliquity and retrograde nature of \myplanet's orbit can allow us to rule out certain migration pathways.
A planet cannot achieve the obliquity of \myplanet by migration through the protoplanetary disc \citep[e.g.,][]{Lin1986OnProtoplanets} and so it likely migrated to its close location after disc dispersal.
Planet-planet interactions (through scattering or von Zeipel-Kozai-Lidov cycles) would similarly struggle to reach the observed obliquity \citep{BeaugeNesvorny2012:PlanetScattering, PetrovichTremaine2016:PlanetPlanet}.
However, von Zeipel-Kozai-Lidov cycles driven by the binary stellar companion could produce the misaligned, and retrograde, orbit of \myplanet \citep{Anderson2016:KL}.\footnote{
See Figure 24 of \cite{Albrecht2022StellarSystems} for a comparison of obliquity distributions produced by various dynamic processes.}

The von Zeipel-Kozai-Lidov mechanism \citep[hereafter vZKL,][]{vonZeipel:1910,Kozai1962SecularEccentricity, Lidov1962TheBodies} suggests that a distant companion can drive alternating cycles of high inclination and eccentricity on the orbit of an inner object and has been widely invoked to explain the close-in orbits of hot Jupiters after subsequent capture by stellar tides \citep[e.g.,][]{Holman1997:16CygniB, WuMurray2003:HD80606, Naoz2011:HJs}.
In particular, \cite{Li2014EccentricitySystems} describes the eccentric von Zeipel-Kozai-Lidov mechanism (EvZKL), where a distant stellar companion on an eccentric, coplanar orbit is capable of flipping an inner orbit into a retrograde sense. 
Their equation 14 presents a compact expression describing the condition for an orbit flip to occur through EvZKL as a function of orbital eccentricities, through which we hope to place some constraints on the migration history of \myplanet.

We note that this condition is not applicable before disc dispersal; the vZKL mechanism applied to a disc \citep{Martin:2014bb} leads to a damping of the inclination down to the critical angle of 39 degrees and so would not be capable of flipping the disc. Despite the torque applied by WASP-94B, disc warping or breaking \citep[e.g. HD100453,][]{Gonzalez:2020bh,Nealon:2020vg} is unlikely due to the typically rapid disc communication timescales in protoplanetary discs \citep{Papaloizou1995OnDisks}.

The orbit flip condition is highly dependent on the (unknown) eccentricity of the stellar binary orbit and the initial semi-major axis of the planet. 
Assuming that the currently observed separation between \mystar and WASP-94B is the semi-major axis of the orbit, the binary orbit would require an eccentricity $> 0.9$ in order to flip a planet orbiting at $a = 10$~au.
At a lower eccentricity ($e \sim 0.6$), a planet with $a = 50$~au could plausibly be flipped. 
For a much more moderate binary eccentricity value of 0.3, the inner planet would need to have formed at hundreds of au from the central star.

The orbits of wide stellar binaries, such as WASP-94, undergo continual evolution under the influence of Galactic tides and stellar flybys \citep{HeggieRasio1996:ClusterBinaries, JiangTremaine2010:WideBinaries, Kaib2013:WideBinary} and could achieve the binary eccentricities required in order to flip the orbit of \myplanet.
However, without further understanding of the binary orbit's eccentricity and modelling of its evolution, it is hard to place constraints on the formation location of \myplanet other than it likely formed at a large distance from \mystar. 
With the above conclusions that \myplanet must have undergone high-eccentricity migration driven by the vZKL mechanism after disc dispersal, we would expect to observe a relatively high C/O ratio compared to planets that had migrated through the disc and accreted inner disc material \citep[e.g.,][]{Madhusudhan2014TowardsMigration,Booth2017ChemicalDrift}.

The WASP-94 system architecture may have led to frequent cometary bombardment. Observations of close-in planets coexistent with debris belts reminiscent of the Solar System's Kuiper belt \citep[e.g.,][]{Plavchan2020:AuMicb} suggest that ready reservoirs of pollutant planetesimals may be common in short period exoplanet systems, even if they are not always observable. 
Further, observations of near-Sun comets in the Solar system \citep{Jones2017:SunComets}, and exo-comet systems with similar orbital periods to \myplanet \citep{Boyajian2016:TabbysStar, Rappaport2018:ExoComets, Zieba2019:betapic, Kiefer2023:ExoComet}, show that cometary bodies are capable of reaching hot Jupiter distances. 
\cite{YoungWyatt2024:TabbysStar} find that the vZKL mechanism can cause an entire planetesimal belt to achieve the extreme eccentricities which would be required to reach the orbital location of \myplanet.
Thus, the dynamic processes which caused the peculiar orbit of \myplanet, could have also disrupted planetesimal reservoirs and encouraged extremely close forays into the planet's orbital region which may have ended with atmospheric enrichment.

The accretion of cometary material could change the atmospheric  C/O ratio and metallicity after the planet has migrated to a close-in location (e.g.\ AF Lep b, \citealt{Zhang2023:AFLepb}).
Although single cometary impact events, such as the 1994 Shoemaker-Levy 9 Jupiter event, may only impart short-term changes such as brightening events \citep{Nicholson1995:SL9Impact} and atmospheric dispersion of deposited material \citep{SanchezLavega1998:SL9Impact}, more sustained bombardment of cometary material may be able to drive larger scale atmospheric changes.
\citet{SainsburyMartinez2024:HJComets} use a 1D atmospheric model coupled with a parametrized comet impact model to investigate the response of hot Jupiter atmospheres to cometary impacts with approximately the same chemical composition as 67P/Churyumov-Gerasimenko \citep{LeRoy2015:Rosetta}. 
For an HD209458~b-like atmosphere with a fiducial C/O ratio of $0.58$, \cite{SainsburyMartinez2024:HJComets} find that continuous bombardment of the atmosphere by comets could drive the atmospheric C/O down to $0.42-0.48$ depending on the mass delivered.
However, such a decrease in C/O ratio would also be expected to coincide with an increase in atmospheric metallicity above stellar values, a feature which is not seen in the atmosphere of \myplanet.

\subsubsection{Inferences from comparing planetary and stellar C/O and metallicity }

\myplanet's sub-stellar atmospheric C/O of $0.49\pm 0.11 = 0.7\pm 0.17 (C/O)_*$ (based on our fiducial, \pRT equilibrium chemistry model using \eureka R=400) may provide some clues into its formation history, given its overall metallicity is close to that of the host star. In particular, if one makes the simplest assumption possible for the composition of the solids and gas from which \myplanet formed, i.e., that the total abundances of the species in the disc add up to the star's composition with some species condensed into dust and others left in the gas phase, then it is not possible to reproduce the planet's metallicity and C/O ratio. Instead, a planet formed out of such a reservoir would necessarily have a stellar C/O ratio if it had a stellar metallicity. As a result, the planet must either have accreted solids and gas from different parts of the disc (where the frozen-out species differ), or the disc's composition must have been modified. Accounting for the loss of some species from the \myplanet's atmosphere due to silicate cloud formation would imply that the planet's bulk C/O ratio is lower than the observed atmospheric composition, only worsening the problem \citep[e.g.,][]{Calamari2024}.

The difference in C/O ratio between the planet and the star could be explained if the planet accreted its gas far from the host star, where the gas phase abundances are low \citep[see, e.g.][]{Zhang2021,Bergin2024}. The C/O ratio of the planet would then reflect the composition of the solids accreted, which generally have a sub-stellar C/O between the water snow line ($\sim 1\,{\rm au}$) and the location where the last significant carbon carrier disappears from the gas ($20-30\,{\rm au}$; see, e.g., \citealt{Oberg2019}). In this scenario, the solids could either be accreted when the planet is still forming if it migrated through the disc\footnote{However, existing planetesimal accretion models do not readily produce such a population \citep[e.g.,][]{Madhusudhan2014TowardsMigration,Madhusudhan2017AtmosphericAccretion,Penzlin2024BOWIE-ALIGN:Compositions}, because gas accretion is typically assumed to continue throughout the disc's lifetime. These studies assumed stars and discs with a solar composition, but conclusions for abundances normalised to the stellar abundances are relatively insensitive to the assumed composition \citep[e.g.,][]{Turrini2021}.}, during the phase of high-eccentricity migration responsible for the planet's high obliquity, or even potentially afterwards due to cometary bombardment.

Formation by pebble accretion could also explain \myplanet's C/O and C/H ratios if the enrichment of the inner disc by ices evaporating from the sublimating pebbles is accounted for \citep[e.g.,][]{Booth2017ChemicalDrift,Booth2019,Schneider2021HowC/O}. In this scenario, \myplanet would need to have accreted its gaseous envelope in the region where CO$_2$ or water are released by evaporation from the pebbles (inside $\sim 5\,{\rm au}$), raising the gas phase C/H abundance back to the stellar value while reducing the C/O ratio. The need for the planet to accrete gas close to its star is however difficult to reconcile with the constraints from its orbital obliquity.


Less volatile elements, such as Na or S, can help determine whether pebble accretion or planetesimal/cometary enrichment scenarios are more likely. In particular, the abundances of these species would likely be lower in the pebble accretion scenario because the atmospheric metals would have been accreted predominantly in gaseous form \citep[e.g.,][]{Chachan2023BreakingGiants,Crossfield2023}.


We assess the possibility of using volatile elements to constrain pebble accretion vs planetesimal/cometary accretion by comparing the measured sodium abundances between the star and the planet. 
\mystar's sodium abundance is Na/H $= 2.63 \pm 0.25 \times$~solar \citep{Teske2016THEB}, assuming a sodium solar abundance from \citet{Asplund2009TheSun} of $\log_{10}\mathrm{Na} = 6.24 \pm 0.04$ sodium atoms for $10^{12}$ hydrogen atoms. 
We compute \myplanet's abundance of sodium relative to hydrogen, Na/H, using the measured VMR of Na in \myplanet's atmosphere from \citet{Ahrer2024AtmosphericHARPS} and the hydrogen in the atmosphere calculated by our best-fitting equilibrium chemistry model. Following \citet{Welbanks2019MassMetallicityK} we find Na/H $= 0.079^{+3.1}_{-0.072} \times$~solar for \myplanet. Note that this is a lower limit as the VMR of Na is relatively unconstrained and the sodium has been retrieved separately from the abundance of hydrogen. 
So \myplanet's Na abundance can be estimated as substellar to stellar at $0.030^{+1.2}_{-0.028}$ times the Na abundance of its host star (likely a lower limit). This suggests a (very) slight preference for the pure pebble accretion scenario which would cause a very low Na/H compared to the star, however, a planetesimal scenario is equally likely as this would allow a stellar to super-stellar Na/H. Further constraints on the Na abundance are needed e.g.\ by gathering more high resolution observations and modelling them together with the low resolution spectrum to get a tighter constraint on the Na abundance. 

Altogether, the orbital obliquity, C/O ratio, metallicity, and Na/H abundance hint at a planet that formed far from its star before reaching its current location by high-eccentricity migration, with \myplanet acquiring its volatiles from planetesimals accreted along the way or subsequent bombardment from outer disc material.

\begin{figure}
    \centering
    \includegraphics[width=\linewidth]{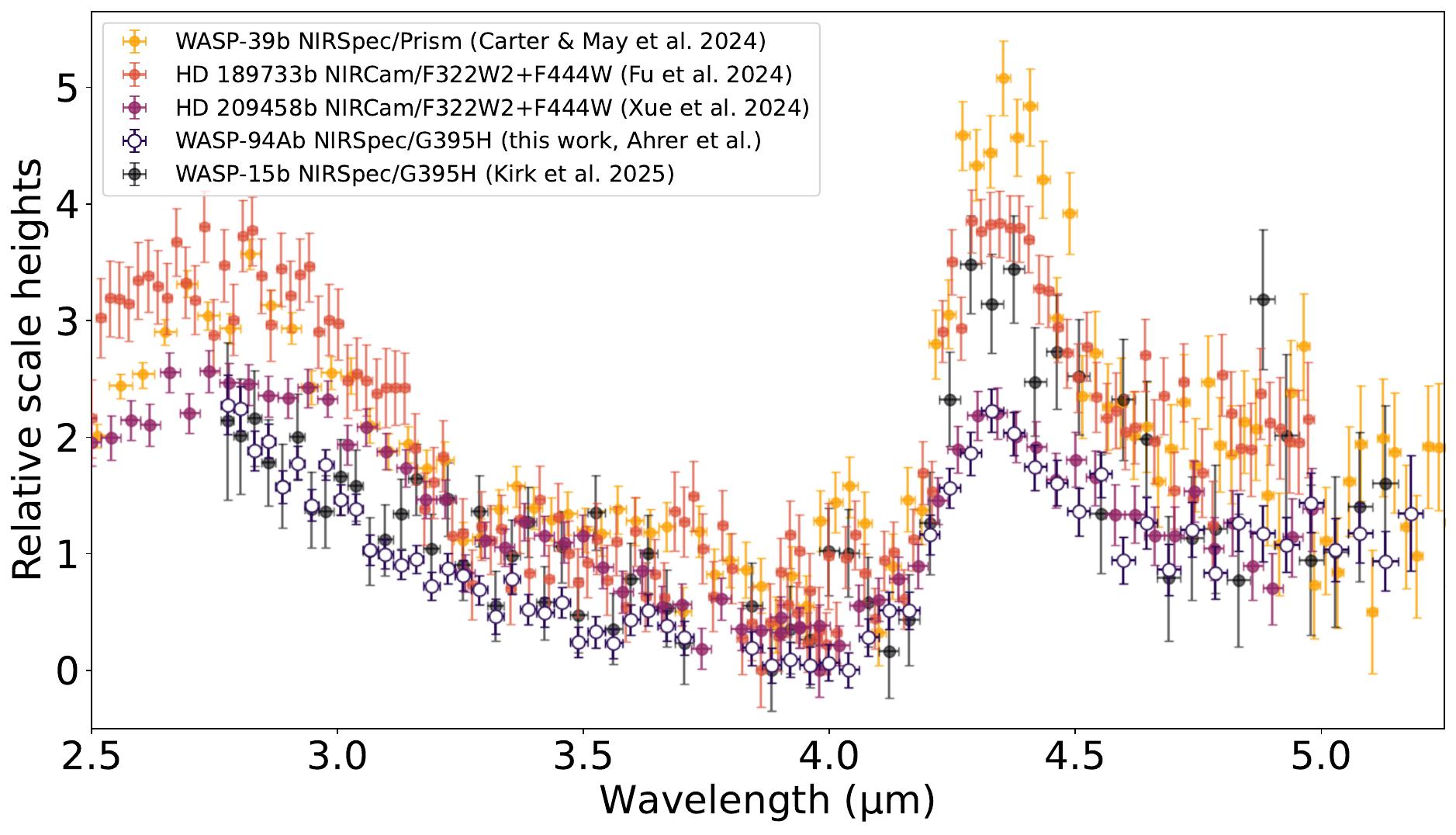}
    \caption{\myplanet in context with JWST transmission spectra published for hot Jupiters (\revision{here assumed as exoplanets with} temperatures $1,000$\,K -- $2,000$\,K, radii R$_\mathrm{p} > 0.5$\,R$_\mathrm{Jup}$, masses M$_\mathrm{p} > 0.2$\,M$_\mathrm{Jup}$) in the 2.5 -- 5.2\,\microns wavelength range: WASP-39b \citepalias{Carter2024AWASP-39b}, HD\,189733b \citep{Fu2024HydrogenExoplanet}, HD\,209458b \citep{Xue20242}, and WASP-15b \citep{Kirk2025WASP-15}. The colours represent the order of temperatures, from light to dark: WASP-39b ($\sim1170$\,K), HD\,189733b ($\sim1200$\,K), HD\,209458b ($\sim1460$\,K),  WASP-94Ab ($\sim1500$\,K; best-fitting offset applied), and WASP-15b ($\sim1680$\,K). The scale heights are computed using a mean molecular mass of $\mu = 2.3$\,amu and normalised using the minimum of the transit depth.  }
    \label{fig:jwst-HJs-transmission-spectra}
\end{figure}

\subsection{JWST transmission spectra of hot Jupiters}

Only a small sample of hot Jupiters (temperatures $1,000$\,K -- $2,000$\,K, radii R$_\mathrm{p} > 0.5$\,R$_\mathrm{Jup}$, masses M$_\mathrm{p} > 0.2$\,M$_\mathrm{Jup}$) have published JWST transmission spectra thus far. In Fig.\,\ref{fig:jwst-HJs-transmission-spectra}, we compare \myplanet's transmission spectrum with the four other hot Jupiters for which spectra in the wavelength range of $2.5 - 5.2$\microns were available. The coolest planet of the sample, WASP-39b \citepalias{Carter2024AWASP-39b}, shows the largest \ce{CO2} absorption feature in atmospheric scale heights in addition to a strong \ce{H2O} and the \ce{SO2} feature ($\sim 4$\microns). The second coolest planet, HD\,189733b shows similarly large features though excluding the \ce{SO2} absorption \citep{Fu2024HydrogenExoplanet}. The hottest out of the sample, WASP-15b \citep{Kirk2025WASP-15}, also shows a large \ce{CO2} bump and a tentative \ce{SO2} detection. This trend also follows the description that higher atmospheric metallicity is needed to produce the photochemical product \ce{SO2}. By eye, \myplanet is most similar to HD\,209458b \citep{Xue20242}, in terms of \ce{CO2}, lack of detection of \ce{SO2} and in temperature, though \myplanet's \ce{H2O} slope is potentially less steep. 

\section{Conclusions}
\label{sec:conclusions}
We presented JWST NIRSpec/G395H transit spectroscopy observations of \myplanet, a hot Jupiter in a retrograde and misaligned orbit around its F-type host star. We used three independent pipelines to determine the planet's transmission spectrum. We also probed for limb asymmetries, but while the morning and evening spectra do appear to show evidence for differences, it is not statistically favoured over a uniform limb approach. 

\myplanet's transmission spectrum shows strong \ce{H2O} and \ce{CO2} absorption bands with detection significances of $4\sigma$ and $11\sigma$ from both our retrievals, \pRT and \hydra. The former relatively low detection significance is likely due to the fact that the full water feature is not covered within G395H's wavelength range (see e.g.\ the comparison to HD\,209458b's spectrum in Fig.\ref{fig:jwst-HJs-transmission-spectra}). When excluding \ce{H2O} both our retrieval setups retrieved high offsets between the NRS1 and NRS2 detectors ($190$\,ppm) and very high \ce{CO2} abundance to account for the non-presence of \ce{H2O} in the model but still managed to get comparatively statistically valid fit. When fixing the detector offset to the value found by the base retrieval, we find higher detection significance for \ce{H2O} at $\sim 11 \sigma$, concluding that the driving factor is the uncertainty in the detector offset. We further find tentative evidence for \ce{CO} (\hydra: $2.8\sigma$, \pRT: $3.3\sigma$) and \ce{H2S} (\hydra: $2.9\sigma$, \pRT: $2.1\sigma$) using our independent retrieval setups. 

We include an offset between the two NIRSpec/G395H detectors NRS1 and NRS2 in our retrieval analysis (Table\,\ref{tab:offsets}) as our reductions show slight offsets in the NRS1 detector and are also preferred based on the larger Bayesian evidence values. We caution the community for future NIRSpec/G395H studies as we find that the inclusion of an offset may significantly change the detection significances for some of the molecules as well as influence the C/O ratio inferred by equilibrium chemistry. A larger sample size (with a range of number of groups/integration) of hot Jupiter observations with NIRSpec/G395H and further investigation by the community is needed to achieve an accurate picture of whether an offset is necessary to be included between NRS1 and NRS2 for all observations and how large it can realistically be. 

Using our best-fit equilibrium chemistry to our fiducial data set (\eureka, R=400), we derive a C/O ratio for \myplanet's atmosphere of $0.49^{+0.08}_{-0.13}$, which is consistent within the uncertainties with the derived C/O for the two other independent data reductions and at both resolutions we computed for this analysis (R=100 and R=400). Similarly consistent is our retrieved atmospheric metallicity of $2.17 \pm 0.65$ $\times$ solar. While \myplanet's metallicity is in line with the stellar metallicity ($2.09 \pm 0.02 \times$ solar), the planet's C/O ratio is $0.70 \pm 0.17 \times$ the star's C/O ratio. 

Formation by pebble accretion is one mechanism that would naturally explain \myplanet's composition assuming evaporation from pebbles in the inner disc, i.e., \ce{CO2} or water is released to the gas and therefore reducing its C/O ratio. Other possibilities include planetesimal accretion in combination with large-distance migration or accretion of solids and gas from different parts of the disc. We find that this lower C/O ratio is unlikely to be caused by the accretion of cometary impact events. While this can decrease the atmospheric C/O, it would also increase the atmospheric metallicity, which is not the case for \myplanet. 
Note that there is the caveat that silicate clouds likely form in \myplanet's atmosphere which we are not sensitive to using NIRSpec/G395H. This could further affect the measured C/O ratio. 

Further observations of \myplanet are necessary to achieve tighter constraints on the molecular abundances and clouds within its atmosphere, which in turn allows better-constrained C/O ratio and metallicity measurements. In addition, in order to rule out possible formation scenarios for \myplanet we require additional elemental abundances, e.g., those of refractory species. Using the Na abundance reported for \myplanet in the literature, we find a sub-stellar to stellar Na/H which can be both explained by pebble and planetesimal accretion. 

Our observations showed that \myplanet is one of the most favourable targets for transmission spectroscopy as its large scale height and the brightness of the star results in uncertainties per scale height comparable to HD\,209458b (see Fig.\,\ref{fig:jwst-HJs-transmission-spectra}).  With our observations we placed meaningful constraints on the C/O ratio and metallicity of \myplanet based on the molecular detections using JWST NIRSpec/G395H, This points the way to an even more complete characterisation and stronger constraints on formation and evolution with complementary data sets.

\section*{Acknowledgements}

This work is based on observations made with the NASA/ESA/CSA JWST. The data were obtained from the Mikulski Archive for Space Telescopes at the Space Telescope Science Institute, which is operated by the Association of Universities for Research in Astronomy, Inc., under NASA contract NAS 5-03127 for JWST. These observations are associated with program \#3154. 
EA is grateful to L. Welbanks for providing details about the Na/H calculations following \citep{Welbanks2019MassMetallicityK} and P. Molli\`{e}re for \pRT support. JK acknowledges financial support from Imperial College London through an Imperial College Research Fellowship grant. Support for program \#3154 was provided to JT by NASA through a grant from the Space Telescope Science Institute, which is operated by the Association of Universities for Research in Astronomy, Inc., under NASA contract NAS5-03127.
CHM gratefully acknowledges the Leverhulme Centre for Life in the Universe at the University of Cambridge for support through Joint Collaborations Research Project Grant GAG/382. RN acknowledges funding from UKRI/EPSRC through a Stephen Hawking Fellowship (EP/T017287/1). PJW acknowledges support from STFC through consolidated grant ST/X001121/1. MZ acknowledges support from a UKRI Future Leaders Fellowship [Grant MR/T040866/1], a Science and Technology Facilities Funding Council Nucleus Award [Grant ST/T000082/1], and the Leverhulme Trust through a research project grant [RPG-2020-82]. VP acknowledges support from the UKRI Future Leaders Fellowship grant (MR/S035214/1, MR/Y011759/1) and UKRI Science and Technology Facilities Council (STFC) through the consolidated grant ST/X001121/1.
We thank the anonymous referee for their time and valuable feedback which improved the paper. 

\section*{Data Availability}
The data products (light curves, transmission spectra, retrieval products) associated with this manuscript are available via Zenodo at  \url{https://doi.org/10.5281/zenodo.15430203}.  The data can be downloaded from the  Mikulski Archive for Space Telescopes (MAST) using the JWST GO programme ID \#3154 or via \url{http://dx.doi.org/10.17909/gsq8-zc73}.
 


\input{references.bbl}
\bibliographystyle{mnras}




\appendix

\section{\revision{Asymmetric limb study for all reductions with \texttt{catwoman}}}
\label{sec:appendix_catwoman-all-reductions}
\begin{figure}
    \centering
    \includegraphics[width=\linewidth]{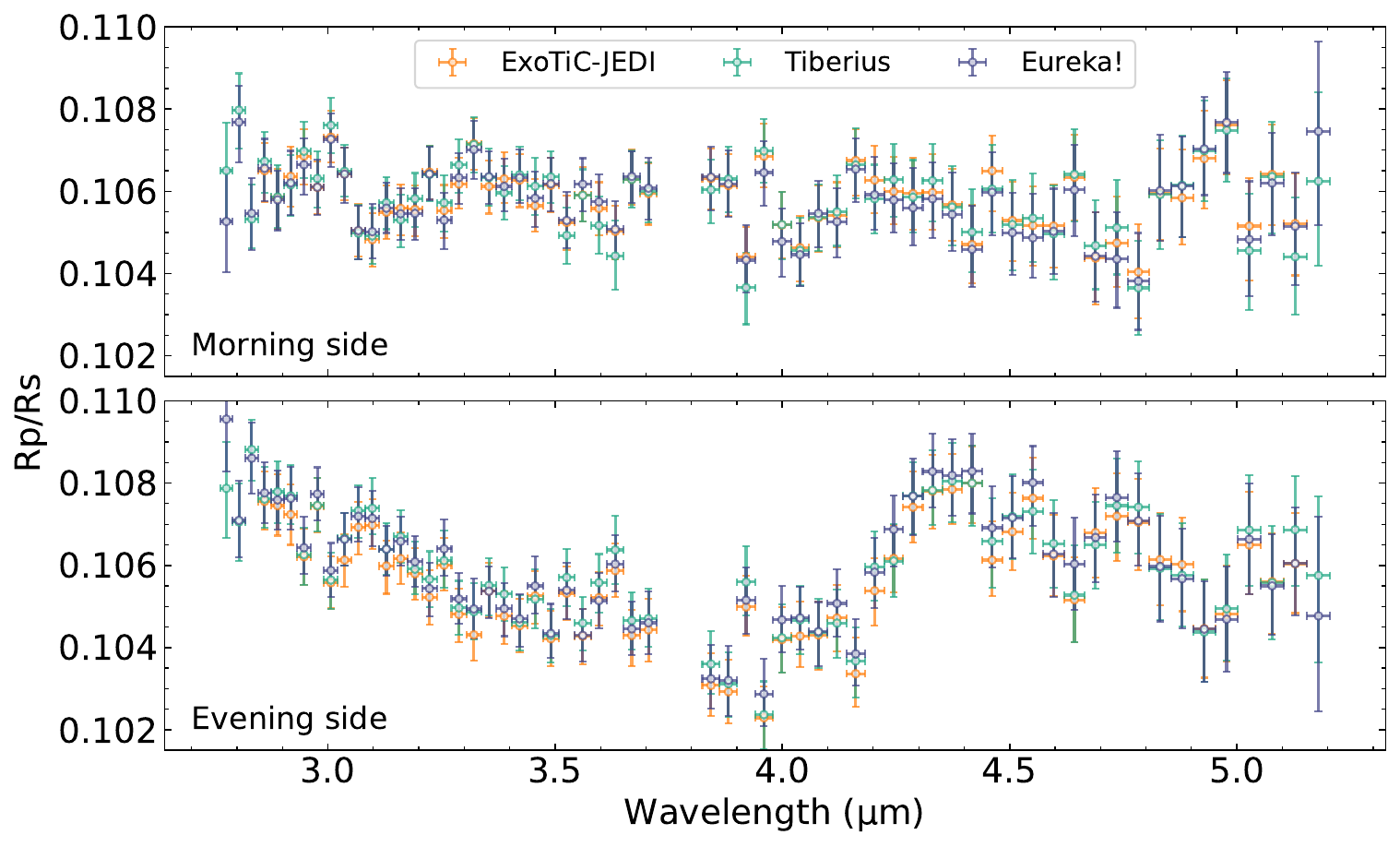}
    \caption{\revision{Morning (top) and evening (bottom) transmission spectra of \myplanet using the R=100 light curves for the three independent reductions, \eureka (dark blue), \tiberius (turquoise) and \exoticjedi (orange).}}
    \label{fig:all-morning-evening}
\end{figure}

\revision{Following the result from applying an asymmetric transit model using \texttt{catwoman} \citep{Jones2021Catwoman:Curves,Espinoza2021ConstrainingSpectroscopy} to the \eureka R=100 light curves in Section\,\ref{sec:trans_spec}, we demonstrate here that with the same setup applied to the R=100 light curves from the \exoticjedi and \tiberius reductions we obtain consistent results, see Fig.\,\ref{fig:all-morning-evening}. For all reductions and light curves, the difference in Bayesian evidence between the simple transit model and the asymmetric model is not statistically significant with differences in Bayesian evidences $\Delta \ln \mathcal{Z} < 1.5$. 
}

\section{Retrieval prior ranges}
\label{sec:appendix_priors}

Here we provide tables with the prior ranges of our retrieval setups, \hydra and \pRT (Table\,\ref{tab:ret_priors}) and the references for the line lists used in our retrievals. 

\begin{table}
    \centering
    \caption{Parameters and uniform prior ranges for our retrieval of \myplanet with \hydra (top) and \pRT equilibrium and free chemistry retrievals (\revision{bottom}; chemical setup is highlighted in bold). Note that they are all uniform except the planetary mass parameter. The rightmost column refers to the literature references of the species as follows: [1] \citet{Polyansky2018ExoMolWater}, [2] \citet{Rothman2010HITEMPDatabase}, [3] \citet{Coles2019ExoMolAmmonia},[4] \citet{harris2006-hcn, Barber2014ExoMolHNC}, [5] \citet{Chubb2020ExoMolAcetylene}, [6] \citet{Azzam2016ExoMolH2S,chubb2018-h2s}, [7] \citet{Yurchenko2020ExoMolCO2}.}
\label{tab:ret_priors}
    \begin{tabular}{lccc}
    \hline
    \multicolumn{4}{c}{\hydra} \\ \hline
 & \textbf{Parameter}              & \textbf{Prior Range}  & \textbf{Reference}\\
\hline
Free   & $\log(\mathrm{H_2O})$  & -15 $\rightarrow$ -0 & [1]\\
 chemistry           & $\log(\mathrm{CO})$ & -15 $\rightarrow$ -1 & [2]\\
(VMR)            & $\log(\mathrm{CO_2})$ & -15 $\rightarrow$ -1 & [2]\\
            & $\log(\mathrm{CH_4})$ & -15 $\rightarrow$ -1 & [2]\\
            & $\log(\mathrm{NH_3})$ & -15 $\rightarrow$ -1 & [3]\\
            & $\log(\mathrm{HCN})$ & -15 $\rightarrow$ -1 & [4] \\
            & $\log(\mathrm{C_2H_2})$ & -15 $\rightarrow$ -1 & [5]\\
            & $\log(\mathrm{H_2S})$ & -15 $\rightarrow$ -1 & [6]\\
\hline
Temp. Profile & $T_\mathrm{1mbar}$ (K) & 300 $\rightarrow$ 2500 \\
            & $\alpha_1$ (K) & 0 $\rightarrow$ 1 \\
            & $\alpha_2$ (K) & 0 $\rightarrow$ 1 \\
            & $\log(P_1 )(\mathrm{bar})$ & -6 $\rightarrow$ 2 \\
            & $\log(P_2 )( \mathrm{bar})$ & -6 $\rightarrow$ 2 \\
            & $\log(P_3 )( \mathrm{bar})$ & -2 $\rightarrow$ 2 \\
\hline
Ref. Pressure & $\log(P_\mathrm{ref} / \mathrm{bar})$ & -6 $\rightarrow$ 2 \\

\hline
Clouds/hazes & $\log(\alpha_\mathrm{haze})$ & -4 $\rightarrow$ 6\\
             & $\gamma_\mathrm{haze}$ & -20 $\rightarrow$ -1\\
             & $\log(P_\mathrm{cl})(\mathrm{bar})$ & -6 $\rightarrow$ 2\\
             & $\phi_\mathrm{cl}$ & 0 $\rightarrow$ 1\\
\hline
Offset & $\Delta_{1-2} (\mathrm{ppm}$ & -200 $\rightarrow$ 200 \\ \hline
\multicolumn{3}{c}{} \\ 


    \hline
    \multicolumn{4}{c}{\texttt{petitRADTRANS}} \\ \hline
      & \textbf{Parameter}             & \textbf{Prior Range} & \textbf{Reference}\\ \hline

\textbf{Equilibrium}   & $\mathrm{C/O}$  & 0.1 $\rightarrow$  1.5 \\
\textbf{chemistry}  & $\log(\mathrm{Fe/H})$ & -3 $\rightarrow$  2 \\
\hline
\textbf{Free}  & $\log(\mathrm{H_2O})$ & -10 $\rightarrow$  -1e-6 & [2] \\
\textbf{chemistry}            & $\log(\mathrm{CO_2})$ & -10 $\rightarrow$  -1e-6 
 & [7]\\
(mass fractions)            & $\log(\mathrm{CO})$ & -10 $\rightarrow$  -1e-6 & [2]\\
& $\log(\mathrm{CH_4})$ & -10 $\rightarrow$  -1e-6 & [2]\\
            & $\log(\mathrm{NH_3})$ & -10 $\rightarrow$ -1e-6 & [3]\\
            & $\log(\mathrm{HCN})$ & -10 $\rightarrow$  -1e-6 & [4]\\
            & $\log(\mathrm{C_2H_2})$ & -10 $\rightarrow$  -1e-6 & [5]\\
            & $\log(\mathrm{H_2S})$ & -10 $\rightarrow$  -1e-6 & [6]\\
\hline

    Temp. Profile & $T$ (K) & 500 $\rightarrow$ 2500 \\
\hline
Ref. Pressure & $\log(P_\mathrm{ref}) ( \mathrm{bar})$ & -6 $\rightarrow$ -2 \\
\hline
Planet mass & $M_\mathrm{p} (M_\mathrm{Jup})$ & $\mathcal{N}( \mu=0.456$, \\
& & $\sigma=0.05)$ \\

\hline
Clouds  & $\log(P_\mathrm{cloud}) (\mathrm{bar})$ & -8 $\rightarrow$ 2\\
\hline
Offset &  $\mathrm{offset} (\mathrm{ppm})$ & -200 $\rightarrow$ 200 \\ \hline

    \end{tabular}
    
\end{table}

\section{Comprehensive retrieval results}
\label{sec:appendix_ret_results}

Table\,\ref{tab:full-retrieval-results} shows the retrieval results of all \pRT and \hydra runs of all three reductions at both computed resolutions (R=100, R=400). 

\begin{table*}

    \caption{Atmospheric retrieval results, with $1\sigma$ uncertainties, except in the case of unconstrained parameters. \newline 
    $^*$ $T_\mathrm{ref}$ corresponds to the temperature of the isothermal temperature profile in the case of \pRT and to the temperature at 1mbar in the case of \hydra. The corresponding detector offsets for these retrievals can be found in Table\,\ref{tab:offsets}. }
    \label{tab:full-retrieval-results}
    \begin{tabular}{l c c c c c c c c c} 
    \hline
     Input spectrum  & $\ln \mathcal{Z}$ & Mass &  $T_\mathrm{ref}$\,(K)$^*$ & $P_\mathrm{ref}$\,(bar)    &  $\log \alpha_\mathrm{haze}$ & $\gamma_\mathrm{haze}$ & $\log P_\mathrm{cloud}$\,(bar) & $\phi_\mathrm{cl}$   \\ 
     \hline
     \eureka, $R=400$   \\
     \pRT: equilibrium chemistry  & $1875.1 \pm 0.1$& $0.472\pm0.049$ & $955 \pm 44$ & $-3.15 \pm 0.31$ & -- & -- &  $-2.60 \pm 0.23$ & -- \\ 
     \pRT: free chemistry & $1871.4 \pm 0.1$& $0.474\pm0.048$ & $702 \pm 82$ & $-3.34 \pm 0.43$ & -- & -- &  $-0.5 \pm 1.6$ & -- \\ 
     \hydra: free chemistry  & $1872.4 \pm 0.1 $& -- &  $814 \pm 76$  &  $-3.58 \pm 0.35$  &  $0.94 \pm 3.00$  &  $-11 \pm 6$  &  $-0.8 \pm 1.8$  &  $0.44 \pm 0.32$  \\ 
    \\ 
    \tiberius, $R=400$ \\
     \pRT: equilibrium chemistry & $1928.1 \pm 0.1$& $0.464\pm0.039$ & $948 \pm 29$ & $-3.05 \pm 0.16$ & -- & -- &  $-2.33 \pm 0.13$ & -- \\ 
     \pRT: free chemistry & $1925.3 \pm 0.1$& $0.469\pm0.043$ & $648 \pm 61$ & $-3.01 \pm 0.38$ & -- & -- &  $0.1 \pm 1.4$ & -- \\ 
     \hydra: free chemistry & $1924.5 \pm 0.1 $& -- &  $772 \pm 59$  &  $-3.20 \pm 0.28$  &  $0.75 \pm 3.09$  &  $-11 \pm 6$  &  $-0.6 \pm 1.6$  &  $0.43 \pm 0.32$  \\ 
    \\
     \exoticjedi, $R=400$\\
     \pRT: equilibrium chemistry & $1792.8 \pm 0.1$& $0.467\pm0.038$ & $973 \pm 27$ & $-3.11 \pm 0.15$ & -- & -- &  $-2.31 \pm 0.13$ & -- \\ 
     \pRT: free chemistry & $1788.4 \pm 0.1$& $0.482\pm0.045$ & $753 \pm 84$ & $-3.49 \pm 0.52$ & -- & -- &  $0.1 \pm 1.3$ & -- \\ 
     \hydra: free chemistry & $1790.5 \pm 0.1 $& -- &  $828 \pm 66$  &  $-3.41 \pm 0.37$  &  $0.72 \pm 3.03$  &  $-11 \pm 6$  &  $-0.5 \pm 1.6$  &  $0.44 \pm 0.31$  \\ 
     \\
         \eureka, $R=100$ \\
    \pRT: equilibrium chemistry  & $514.5 \pm 0.1$& $0.476\pm0.049$ & $984 \pm 44$ & $-3.25 \pm 0.33$ & -- & -- &  $-2.61 \pm 0.24$ & -- \\ 
     \pRT: free chemistry  & $511.8 \pm 0.1$& $0.476\pm0.046$ & $686 \pm 58$ & $-3.18 \pm 0.47$ & -- & -- &  $0.0 \pm 1.4$ & -- \\ 
     \hydra: free chemistry  & $516.1 \pm 0.1 $& -- &  $779 \pm 51$  &  $-2.98 \pm 0.41$  &  $0.56 \pm 3.12$  &  $-11 \pm 6$  &  $-0.4 \pm 1.5$  &  $0.45 \pm 0.33$  \\ 
    \\
         \tiberius, $R=100$  \\
     \pRT: equilibrium chemistry  & $518.0 \pm 0.1$& $0.462\pm0.045$ & $939 \pm 33$ & $-3.09 \pm 0.21$ & -- & -- &  $-2.36 \pm 0.20$ & -- \\ 

     \pRT: free chemistry & $516.4 \pm 0.1$& $0.461\pm0.043$ & $635 \pm 60$ & $-2.94 \pm 0.39$ & -- & -- &  $-0.3 \pm 1.5$ & -- \\ 
    
     \hydra: free chemistry  & $518.1 \pm 0.1 $& -- &  $745 \pm 52$  &  $-3.05 \pm 0.33$  &  $0.77 \pm 3.08$  &  $-11 \pm 6$  &  $-0.5 \pm 1.5$  &  $0.45 \pm 0.32$  \\ 
    \\
    \exoticjedi, $R=100$ \\
     \pRT: equilibrium chemistry & $480.5 \pm 0.1$& $0.462\pm0.045$ & $973 \pm 30$ & $-3.14 \pm 0.21$ & -- & -- &  $-2.35 \pm 0.19$ & -- \\ 
     \pRT: free chemistry & $477.5 \pm 0.1$& $0.472\pm0.041$ & $675 \pm 63$ & $-3.09 \pm 0.48$ & -- & -- &  $0.2 \pm 1.3$ & -- \\ 
      \hydra: free chemistry   & $480.7 \pm 0.1 $& -- &  $753 \pm 55$  &  $-2.75 \pm 0.45$  &  $0.75 \pm 3.11$  &  $-11 \pm 6$  &  $-0.3 \pm 1.5$  &  $0.44 \pm 0.32$\\  
     \hline
      
    \end{tabular}

\end{table*}

\section{Corner plots}
\label{sec:appendix_corner_plots}
\begin{figure*}
    \includegraphics[width=\textwidth]{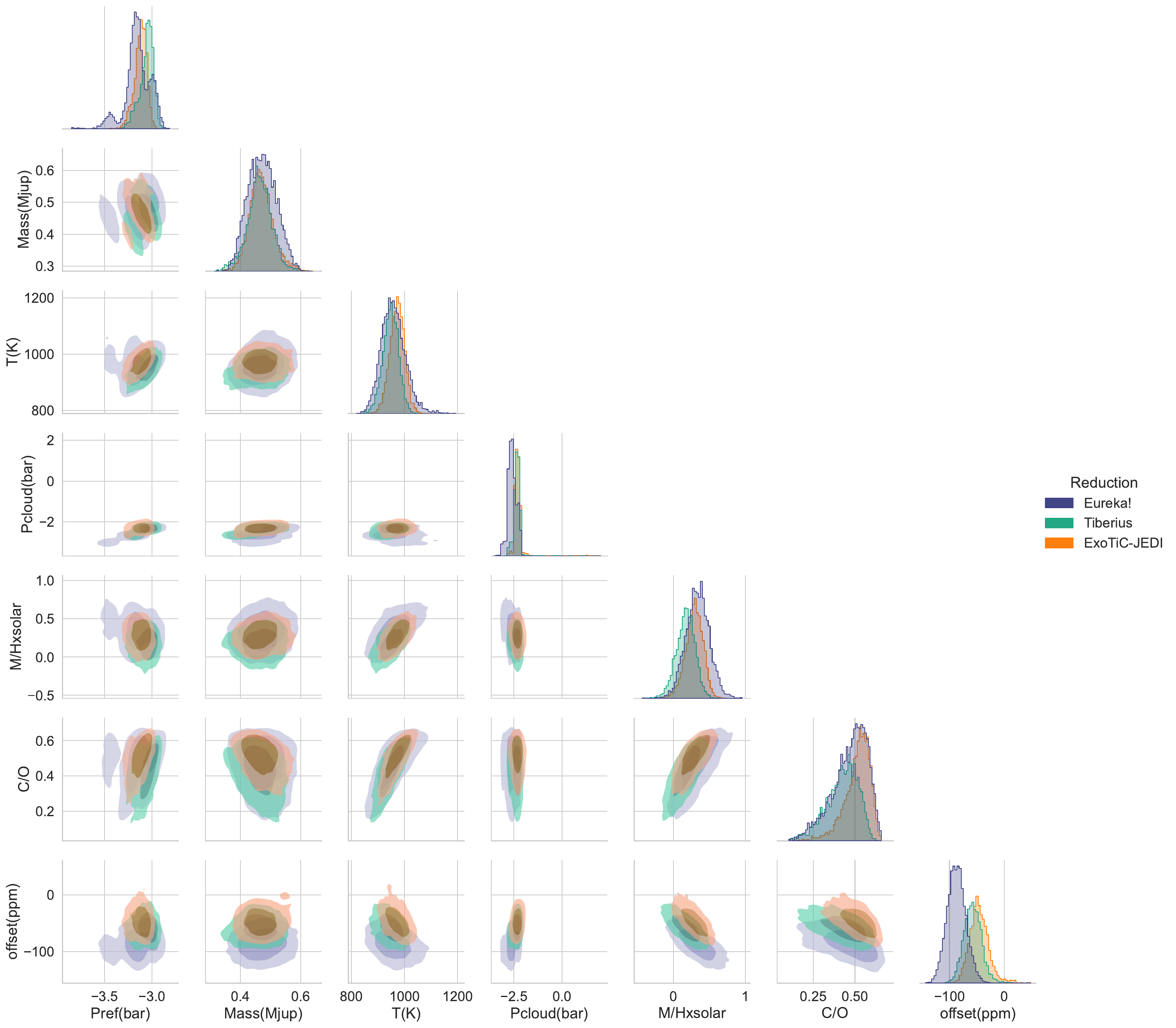}
    \caption{Posterior plots of the \pRT equilibrium chemistry runs for \eureka (purple), \tiberius (turquoise) and \exoticjedi (orange) at R=400. The two shadings correspond to the $1\sigma$ (darker) and $2\sigma$ (lighter) confidence intervals. }
    \label{fig:eq_chem_corner}
\end{figure*}

\begin{figure*}
    \includegraphics[width=\textwidth]{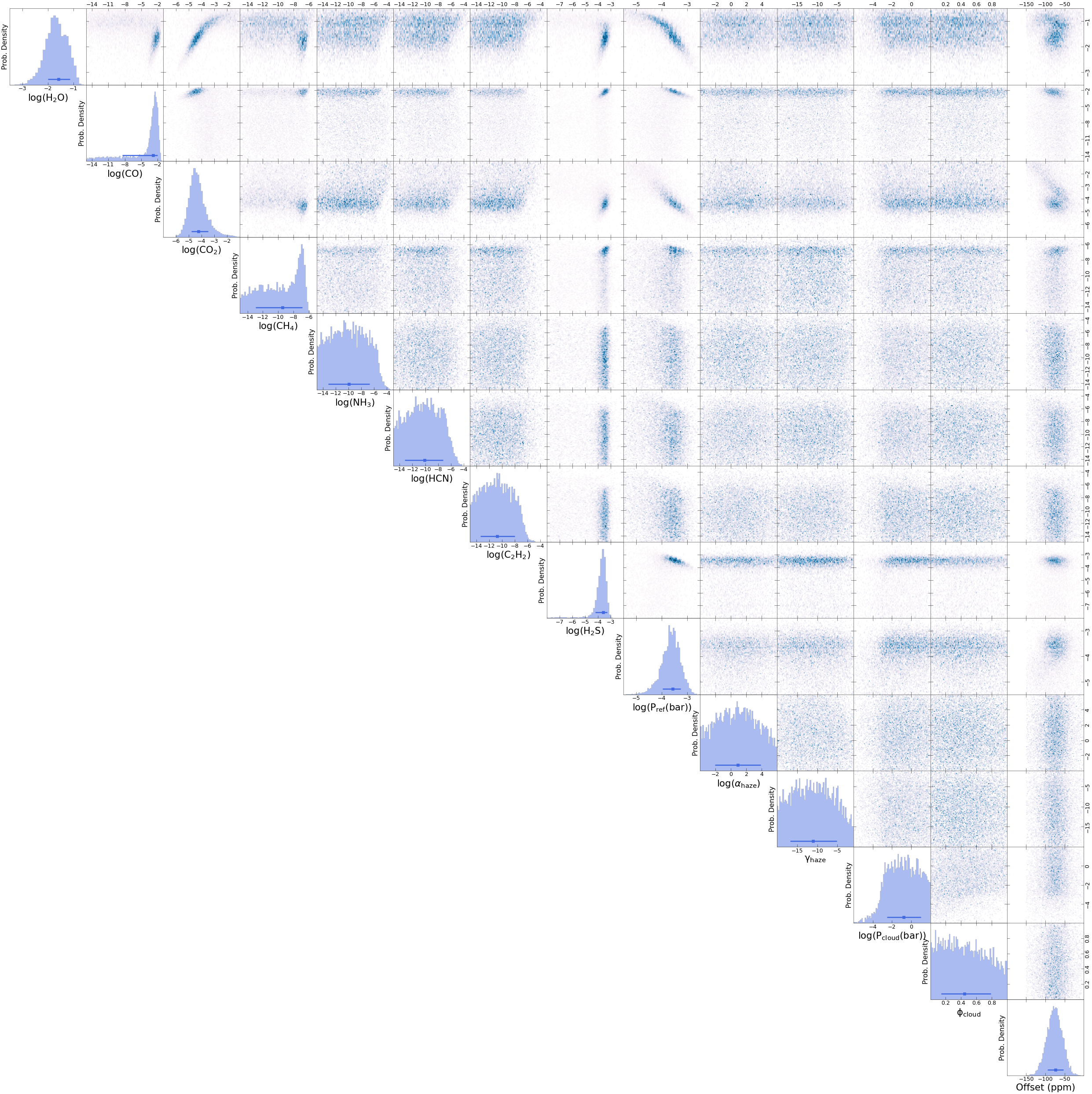}
    \caption{Posterior of the retrieval of \myplanet using the \eureka R=400 dataset using \hydra. We have left out the constraints on the P-T profile parameters for clarity. The constrained temperature profile is given in Figure~\ref{fig:ret}.}
    \label{fig:hydra_corner}
\end{figure*}

Here we include the posterior plots for the \pRT equilibrium chemistry retrieval runs for all reductions in Fig.\,\ref{fig:eq_chem_corner} at R=400, as well as the posterior of the free chemistry \hydra retrieval of \myplanet using the \eureka R=400 dataset in Fig.\,\ref{fig:hydra_corner}.

\section{Equivalent width fitting to unblended O I, C I, and CH lines}
\label{sec:appendix_O_C_lines}

\cite{Teske2016THEB} reported the equivalent widths (EWs) of unblended carbon and oxygen lines measured from Magellan II/MIKE spectra of WASP-94\,A and B (see their extended Table 1), but did not report the absolute abundances derived from these measurements for the individual stars. To translate the WASP-94\,A EWs (specifically the EWA-IR EWs) into abundances, we used the same methodology as described in that paper, via the curve-of-growth analysis within MOOG and the python wrapper \texttt{Qoyllur-quipu} ($q^{2}$; first described in \citealt{Ramirez2014}). We included seven C I lines, three CH lines, and the O triplet lines (with NLTE corrections from \citealt{Ramirez2007} that are built into $q^2$). We used a MARCS 1D-LTE stellar model with the stellar parameters and associated errors listed in Table 1 of this paper (from the Teske et al. WASP-94B reference, isochrone log $g$ analysis). To then calculate [C/H] and [O/H] values for WASP-94A, we assumed the absolute C and O solar abundances from \cite{Asplund2021} and take the logarithm of the difference, that is, $10^{\rm{(WASP94A-solar)}}$. Different combinations of C and O abundance indicators from both these EW measurements and the synthesis fitting are averaged together to get an estimate of 0.68$\pm$0.10 for the C/O ratio of WASP-94\,A. This analysis can be reproduced using the EWs and stellar parameters given in \cite{Teske2016THEB} using $q^2$.  



\bsp	
\label{lastpage}
\end{document}